\documentclass[%
 aip,
 jmp,%
 amsmath, amssymb,
 reprint,%
]{revtex4-2}
\usepackage[top=1in, bottom=1in, left=1in, right=1in]{geometry}

\usepackage[utf8]{inputenc}
\usepackage[T1]{fontenc}
\usepackage{graphicx}
\usepackage{dcolumn}
\usepackage{bm}
\usepackage{hyperref}

\begin{document}


\title{Strong-field Driven Sub-cycle Band Structure Modulation and Dephasing Control}


 \author{Francis Walz$^{\dagger}$}
 \affiliation{Department of Physics and Astronomy, Purdue University, 525 Northwestern Ave., West Lafayette, 47907, Indiana, USA}
 
\author{Shashank Kumar$^{\dagger}$}%
\email{kumar414@purdue.edu}
 \affiliation{Department of Physics and Astronomy, Purdue University, 525 Northwestern Ave., West Lafayette, 47907, Indiana, USA}

\author{Amirali Sharifi Olounabadi}
\affiliation{Department of Physics and Astronomy, Purdue University, 525 Northwestern Ave., West Lafayette, 47907, Indiana, USA}

 \author{Yuyan Zhong}%
 \affiliation{Department of Physics and Astronomy, Purdue University, 525 Northwestern Ave., West Lafayette, 47907, Indiana, USA}

\author{Russell Zimmerman}%
 \affiliation{Department of Physics and Astronomy, Purdue University, 525 Northwestern Ave., West Lafayette, 47907, Indiana, USA}

\author{Siddhant Pandey}%
 \affiliation{Department of Physics and Astronomy, Purdue University, 525 Northwestern Ave., West Lafayette, 47907, Indiana, USA}

 \author{Eric Liu}%
 \affiliation{Department of Physics and Astronomy, Purdue University, 525 Northwestern Ave., West Lafayette, 47907, Indiana, USA}

 \author{Liang Z. Tan}%
 \affiliation{Molecular Foundry, Lawrence Berkeley National Laboratory, 1 Cyclotron Rd., Berkeley, 94720, California, USA}

 \author{Niranjan Shivaram}%
 \email{niranjan@purdue.edu}
 \affiliation{Department of Physics and Astronomy, Purdue University, 525 Northwestern Ave., West Lafayette, 47907, Indiana, USA}
 \affiliation{Purdue Quantum Science and Engineering Institute, 1205 W State St, West Lafayette, Indiana 47907, USA}


\begin{abstract}
Over the past decade, ultrafast electron dynamics in the solid state have been extensively studied using various strong light-matter interaction techniques, such as high-harmonic generation. These studies lead to multiple interpretations of light-matter interaction in the strong-field regime, with exact mechanisms not yet fully understood. It is well known that strong-field interaction with a crystalline solid leads to significant modification of its band structure and, hence, its optical properties on ultrafast timescales. In this work, we present measurements of ultrafast electric-field observables in magnesium oxide using a non-resonant nonlinear optical interaction. Using field observables, we show that strong laser fields modulate the band structure on sub-cycle timescales, thereby altering the material's nonlinear optical response. We perform time-dependent perturbation theory calculations using a field-dependent dispersion relation and non-perturbative semiconductor Bloch equation calculations, both of which agree with experimental observations. Furthermore, we directly extract dephasing times from the real-time signal electric field envelope and show sub-cycle control of dephasing times. Our work offers a new perspective on strong-field-driven electron dynamics in solids through electric-field observables. The demonstrated attosecond modulation of the nonlinear response could have important implications for quantum light generation and quantum spectroscopy using nonlinear optical processes.
\end{abstract}


\maketitle

\noindent $^{\dagger}$These authors contributed equally to this work.

\section{Introduction}\label{sec1}

The interaction of strong, ultrafast laser pulses with solids can give rise to many intriguing phenomena that have been extensively studied using multiple ultrafast techniques. Examples include high-harmonic generation (HHG) \cite{Ghimire2011, Ghimire2019}, the realization of the Haldane model in a laser-dressed crystal \cite{Mitra2024}, and transient optical response modulation, also known as the dynamical Franz-Keldysh effect \cite{Jauho1996, Nordstrom1998, lucchini2016, Reislohner2023}. These effects are fundamental to understanding electron dynamics in the strong field regime. Several theoretical models have been developed to study these phenomena, such as the Houston basis formalism \cite{Vampa2014, Wu2015} and Wannier functions \cite{PhysRevX.7.021017, PhysRevB.100.195201}, and the Semiconductor Bloch equation \cite{Schubert2014, Hohenleutner2015, Huttner2017} to model HHG in solids. To understand the dynamical Franz-Keldysh effect, more sophisticated time-dependent density functional theory (TDDFT) techniques have been used \cite{lucchini2016, Otobe2016}. Despite differences in approach, these models inherently incorporate non-perturbative strong-field effects in the material to gain insight into ultrafast electron dynamics, as measured by various experimental observables. 

In the study presented here, we investigate the effects of electron interactions with strong fields in solids using ultrafast nonlinear optical spectroscopy, specifically electric-field observables in four-wave mixing (FWM). This nonlinear spectroscopy technique involves three coherent femtosecond pulses interacting with a crystalline solid to produce a fourth pulse that is proportional to the material's third-order susceptibility ($\chi^{(3)}$) \cite{Boyd2008}. Previously, the FWM process has been studied in quantum well and quantum dot systems experimentally \cite{10.1063/1.93802, PhysRevB.57.8774} and theoretically \cite{PhysRevB.67.035329, Flayyih2013}. However, in an optically thick semiconductor, the propagation effect of the FWM output pulse becomes complex \cite{10.1063/1.371183, PhysRevB.51.10601, PhysRevB.48.9426}. Recently, with advances in Free-Electron Lasers (FELs), the FWM process using purely extreme ultraviolet (XUV) pulses has also been demonstrated \cite{Bencivenga2015, Foglia2018}. On the other hand, nonlinear response functions have also been measured using HHG in solids \cite{Han2019}. 

In this article, we report an experiment in which we measure electric-field observables for degenerate four-wave mixing (DFWM) in magnesium oxide (MgO) to study the strong-field modification of the band structure and, hence, the nonlinear response function in the material. Using the TADPOLE technique \cite{fittinghoff1996}, the amplitude and phase of the DFWM output field are measured, providing complete information about ultrafast nonlinear light-matter interactions.  We also develop a theoretical model based on time-dependent perturbation theory, combined with a field-dependent dispersion relation \cite{gruzdev2018}, to simulate the DFWM process in a large-band-gap crystal (such as MgO). We study the modulation of the amplitude and the temporal chirp of the output signal as we vary the time delay between incident pulses with attosecond resolution. It was found that both the amplitude and the chirp oscillate with a time delay, and our model, with good agreement, predicts that this oscillation can occur only due to ultrafast modulation of the band structure of MgO under strong electric fields. Additionally, we directly extract dephasing times from our measured signal electric field and show sub-cycle control of dephasing time due to the strong field driven modulation of the band structure.


\begin{figure}[t]
\centering
\includegraphics[width=1.0\textwidth]{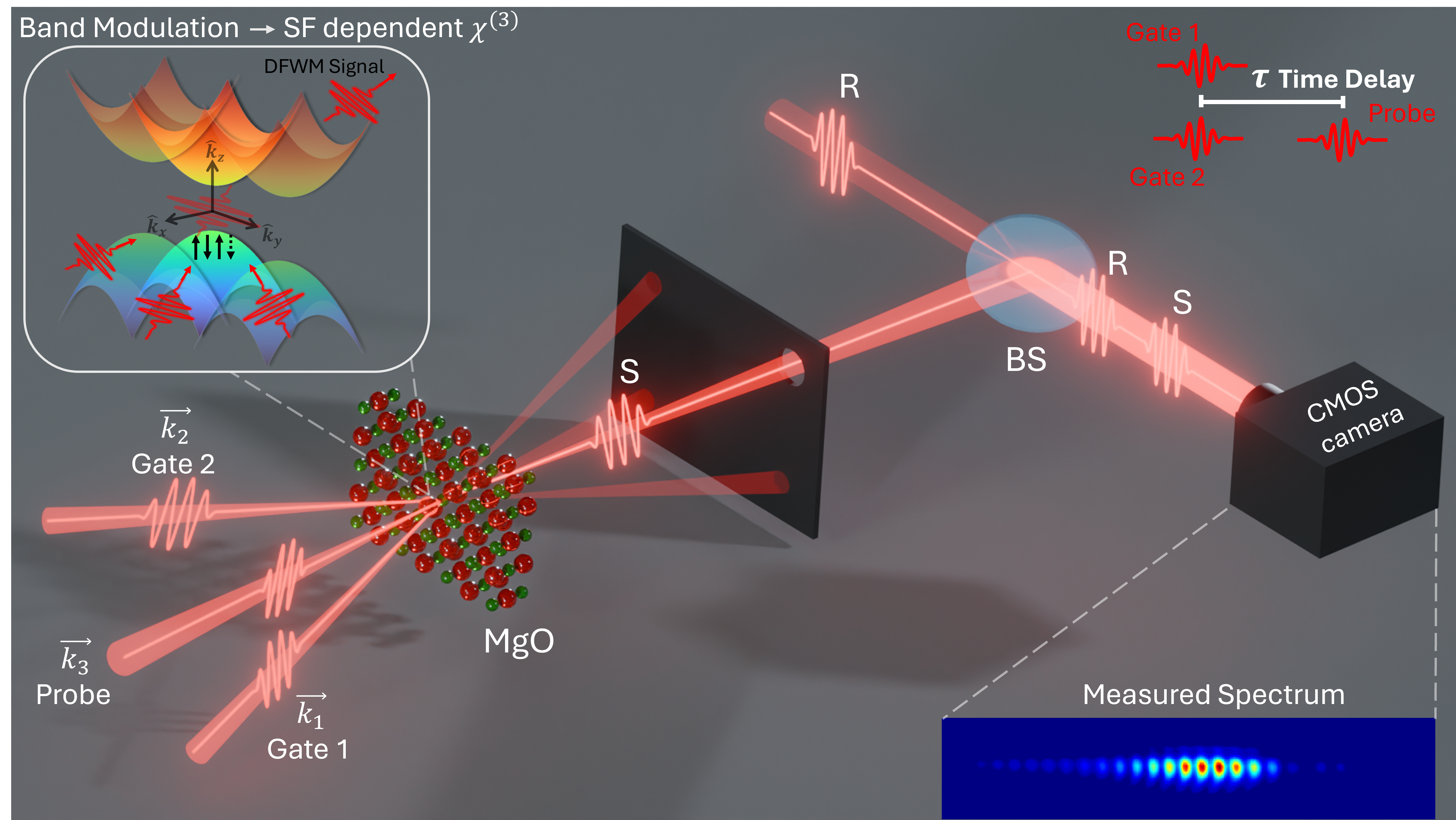}
\caption{(a) \textbf{Ultrafast Band modulation}: The net electric field resulting from the superposition of three ultrafast femtosecond pulses interacts with MgO and significantly modifiesthe band structure of the material. This interaction modulates the band structure (represented by opaque curves) and shifts it (transparent curves) from its original position on an ultrafast timescale. The measured output electric field from the nonlinear process can capture this modulation, providing insights into the dynamics of strong-field interactions in a solid. (b) \textbf{Schematic diagram of the spectral interferometry measurement for degenerate four-wave mixing (DFWM)}: Experimental setup illustrates the interaction of three ultrafast pulses (Gate 1, Gate 2, and Probe) with the magnesium oxide crystal and the generation of a DFWM signal pulse. This signal interferes with a reference pulse at a spectrometer, resulting in a spectral interference measurement. R - Reference pulse, S - Signal pulse, and BS - beam splitter.}
\label{fig1:Expt}
\end{figure}

\begin{figure}[t]
    \centering
    \includegraphics[width=1.0\linewidth]{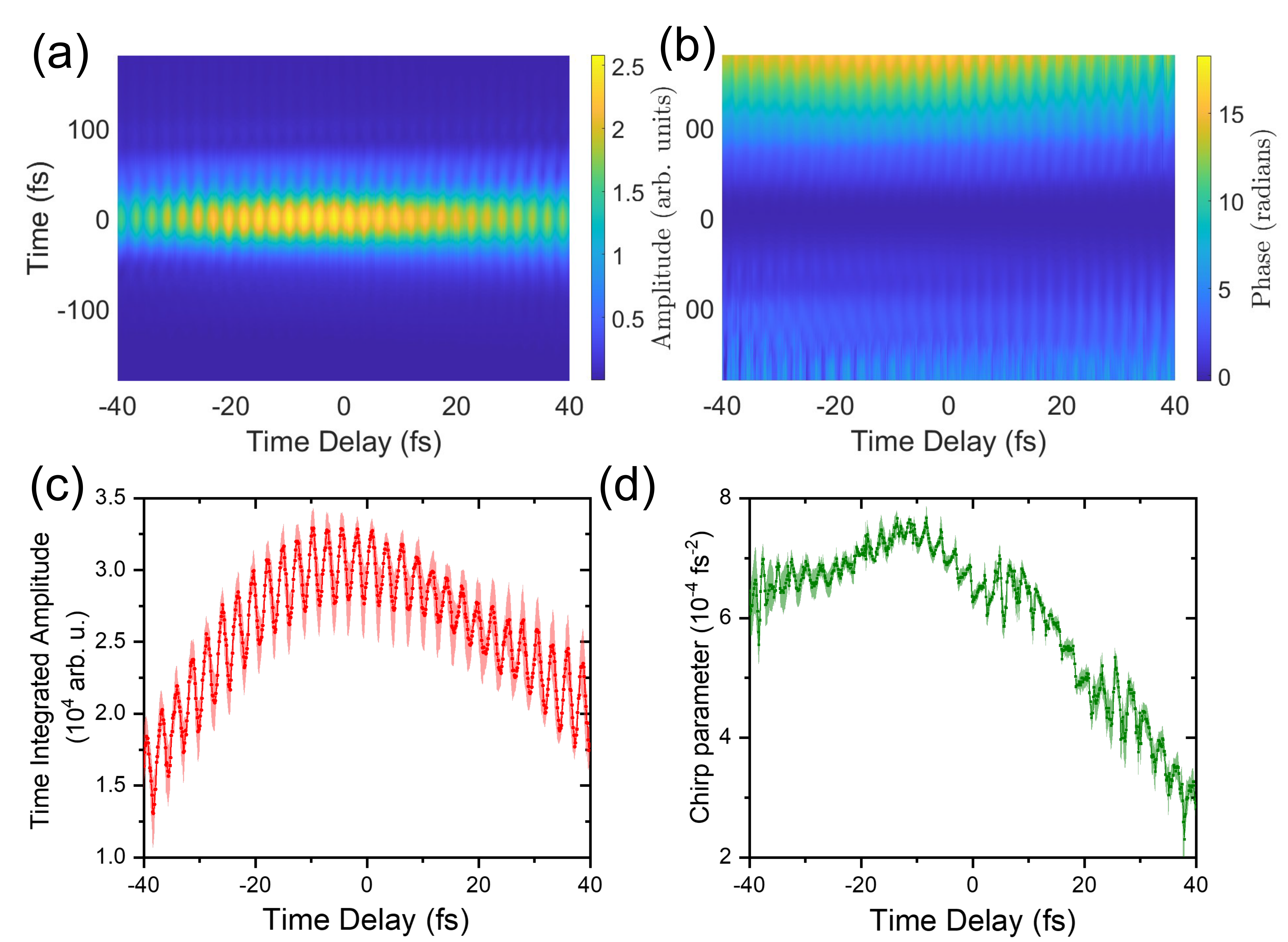}
    \caption{\textbf{Experimentally Measured Amplitude and Phase}: (a) Amplitude $|E_0(t,\tau)|$ and (b) Phase $\varphi(t,\tau)$ of the Degenerate Four-Wave Mixing (DFWM) electric field signal in Magnesium Oxide (MgO). (c) Time integrated temporal amplitude and (d) Temporal Chirp as a function of Time Delay $\tau$ with a $\tau$ delay step of 200 as extracted from (a) and (b), respectively. The shaded region represents the statistical error in the measurement.}
    \label{fig: exp_intensity_chirp}
\end{figure}

\section{Results and Discussion}
\subsection{Amplitude and phase of the measured DFWM electric field}\label{results}
Our experiment employs the spectral interferometry-based technique of TADPOLE \cite{fittinghoff1996} to completely characterize the degenerate Four-Wave Mixing (DFWM) signal generated using three 60 fs near-infrared (NIR) pulses centered around 800 nm ($\hbar \omega_0 \approx 1.56eV$). These three linearly polarized pulses interact with the Magnesium Oxide (MgO) crystal with an intensity of $\sim0.1$ TW/cm$^{2}$ in a BOXCAR configuration \cite{Shirley1980}, producing a fourth beam in a new direction that satisfies the phase matching condition $\bm{k_s} = \bm{k}_1-\bm{k}_2+\bm{k}_3$. Figure~\ref{fig1:Expt} (a) illustrates the schematic representation of the highest valence band and the lowest conduction band of MgO, along with the Four-Wave Mixing (FWM) process occurring within the crystal. This FWM process occurs with a simultaneous modulation of the band by the net electric field present during the interaction, as discussed later. Since the MgO bandgap is about 7.8 eV and the photon energy of each incoming pulse is 1.56 eV, the interaction is highly off-resonant. A reference beam then interferes with the output signal, and spectral interference is measured using a spectrometer. Representative interference fringes from the measurement along with the experimental setup are shown in Figure~\ref{fig1:Expt} (b). From this interference, we can extract the temporal amplitude and phase of the electric field of the DFWM signal (more details in Section~\ref{exp_methods}). The electric field is then measured for each time delay ($\tau$) between two phase-locked Gate pulses and the Probe pulse, with a time delay resolution of 200 as. Therefore, the complete electric field of the DFWM signal can be expressed as $E(t,\tau) = |E_0(t,\tau)|\exp{\left(-i\varphi(t,\tau)\right)}$, where the temporal amplitude $|E_0(t,\tau)|$ and phase $\varphi(t,\tau)$ depend on real time $t$ and the time delay between the Gate and Probe pulses $\tau$. Both the amplitude and the phase are plotted in Figures~\ref{fig: exp_intensity_chirp} (a) and (b), respectively. For each time delay $\tau$, the phase is fit to a fifth-order polynomial,
\begin{equation}
\label{4th_order_poly}
\varphi(t,\tau) = \varphi_{0} + a(\tau)t + b(\tau)t^{2}+c(\tau)t^{3}+d(\tau)t^{4}+e(\tau)t^{5} 
\end{equation}
The second-order coefficient $b(\tau)$, which is also called the temporal chirp, is extracted as a function of the time delay $\tau$. We integrate the extracted amplitude over the total pulse duration for each time delay; this is called the time-integrated amplitude and is defined as $\int |E_0(t,\tau)|\ dt$. The time-integrated amplitude as a function of time delay is shown in Figure~\ref{fig: exp_intensity_chirp} (c). Figure~\ref{fig: exp_intensity_chirp} (d) shows the chirp parameter $b(\tau)$ as a function of the time delay with attosecond resolution. Modulation of nonlinear signals has previously been observed in semiconductors, where a strong near-infrared (NIR) pump pulse generates mid-infrared (MIR) sidebands by mixing with a strong THz pulse \cite{chin2001}. The time delay between the two mixing pulses modulates the spectral intensity of the MIR sideband. However, sub-cycle modulation in electric field observables, as shown here, has not been observed previously.

\subsection{Strong-field Driven Nonlinear Response Modulation}
The output DFWM signal depends on the third-order polarization of the crystal. The spectrum of third-order polarization can be written in terms of the nonlinear response function, i.e., third-order susceptibility $\chi^{(3)}(\omega; \omega, -\omega, \omega)$, which is a material property. However, in strong fields, higher-order contributions can arise. Most generally, the polarization is \cite{Boyd2008},
\begin{equation}\label{P_higher_order}
    P_i(\omega) = \chi^{(3)}_{ijkl}\mathcal{E}_{j}(\omega)\mathcal{E}_{k}(\omega)\mathcal{E}_{l}(\omega) + \chi^{(5)}_{ijklmn}\mathcal{E}_{j}(\omega)\mathcal{E}_{k}(\omega)\mathcal{E}_{l}(\omega)\mathcal{E}_{m}(\omega)\mathcal{E}_{n}(\omega) + \cdots
\end{equation}
If the polarization vectors of the input and output electric fields are in the same direction, as in our experiment, the indices $i,j,k, \cdots$ are the same. For the remainder of this discussion, we omit the polarization subscripts for brevity. Here, $\mathcal{E}(\omega)=(E_1(\omega)e^{i\bm{k_1}\cdot r}+E_2(\omega)e^{i\bm{k_2}\cdot r}+E_3(\omega)e^{i\bm{k_3}\cdot r}+ \mathrm{c.c.})$. Based on the phase-matching condition mentioned above, the only terms that contribute to the measured signal are
\begin{align}\label{no_mod_pol}
    P_{DFWM} = &\chi^{(3)}_0 E_1(\omega)E_2^*(\omega)E_3(\omega)e^{i(\bm{k}_1-\bm{k}_2+\bm{k}_3)\cdot \bm{r}} \notag \\
    &+ \chi^{(5)}_0 E_1(\omega)E_2^*(\omega)E_3(\omega)(|E_1|^2 + |E_2|^2 + |E_3|^2)e^{i(\bm{k}_1-\bm{k}_2+\bm{k}_3)\cdot \bm{r}} + \cdots
\end{align}
Here, $E_1, E_2$ and $E_3$ are the two Gate and Probe pulse electric fields in the frequency domain. The introduction of a time delay $\tau$ between Gate and Probe pulses will result in a phase term of $\exp{(-i\omega\tau)}$ in the electric field of the Gate pulses. However, this phase term does not result in a $\tau$-dependent modulation of the amplitude or phase of the polarization corresponding to the DFWM process when $\chi_0^{(3)}$ is independent of the input fields (for a detailed calculation, see the supplementary material). In the presence of moderately strong fields, nonlinear response functions are expected to be significantly modified as observed, for example, in studies of the dynamical Franz-Keldysh effect using attosecond transient absorption spectroscopy \cite{lucchini2016, chin2001}.  Thus, we model the attosecond modulation of the nonlinear response function due to the strong field interaction of electrons in the crystal such that $\chi^{(3)} \rightarrow \chi^{(3)}(\mathcal{E})$ becomes an instantaneous function of the incoming electric fields. 

 In high-intensity laser fields, excited electron-hole pairs can extend beyond the edge of the Brillouin zone, leading to Landau–Dykhne–type transitions. Such effects have been observed in high-harmonic spectra \cite{uzan2020, uzan2022}. These transitions provide a powerful means to probe modifications of the band structure induced by intense laser pulses. To directly probe band-structure modifications, time- and angle-resolved photoelectron spectroscopy (Tr-ARPES) can be employed, as it images electron dynamics in materials with simultaneous momentum and energy resolution. Ab initio simulations have been performed to study Tr-ARPES in periodic systems under strong fields \cite{Neufeld2022}. Most recently, a Tr-ARPES experiment on bismuth telluride observed field-dressed bands in the presence of an intense laser field \cite{wang2025}. This suggests that the ultrafast modification of the band structure (or adiabatic field-dressed bands) offers a useful framework for understanding off-resonant electron interactions in strong-field regimes. Alternative perspectives for interpreting these field-dressed bands have also been proposed, most notably within the framework of Floquet theory \cite{tiwari2023}.

 For a rigorous quantitative analysis, we start with time-dependent perturbation theory to simulate the DFWM experiment. However, the time-integrated amplitude and the chirp of the calculated signal electric field do not exhibit any of the oscillatory features observed in the experiment (see supplementary material figure \ref{fig:band_vs_noband} red vs blue curves). This aligns with the expectation that the field-free response function accounts only for perturbative wave-mixing processes \cite{Boyd2008, Bruus2004} and cannot include any field-driven modification of the response function. We thus introduce an ultrafast band-modulation model \cite{gruzdev2018} to implicitly account for the strong-field effects on the nonlinear response function. In the presence of an external field, the crystal momentum of the electrons can be written as $\mathbf{p}(t) = \mathbf{p}_0 + e \mathbf{A}_{ext}(t)$ \cite{Vampa2014, keldysh1958effect, keldysh1958influence, Ghimire2019, gruzdev2018, Wu2015}, where $\bm{p}_0$ is the initial momentum of the electron and $\bm{A}_{ext}(t)$ is the vector potential of the external field. This results in the time-dependent modification of the electronic dispersion relation, i.e. $\epsilon(\mathbf{p}) \rightarrow \epsilon(\mathbf{p}_0 + e \mathbf{A}_{ext}(t))$ where $\epsilon$ is the energy of the electron. The details of the calculations are listed in the supplementary material. A theoretical investigation of the nonlinear optical response under strong light-matter interaction in MgO has previously been carried out using the Floquet–Bloch formalism \cite{gorelova2024}. Here, we demonstrate that incorporating non-perturbative effects (specifically, band-structure modulation) within a perturbative framework yields excellent agreement with the experiment.


\begin{figure}[t]
    \centering
    \includegraphics[width=\linewidth]{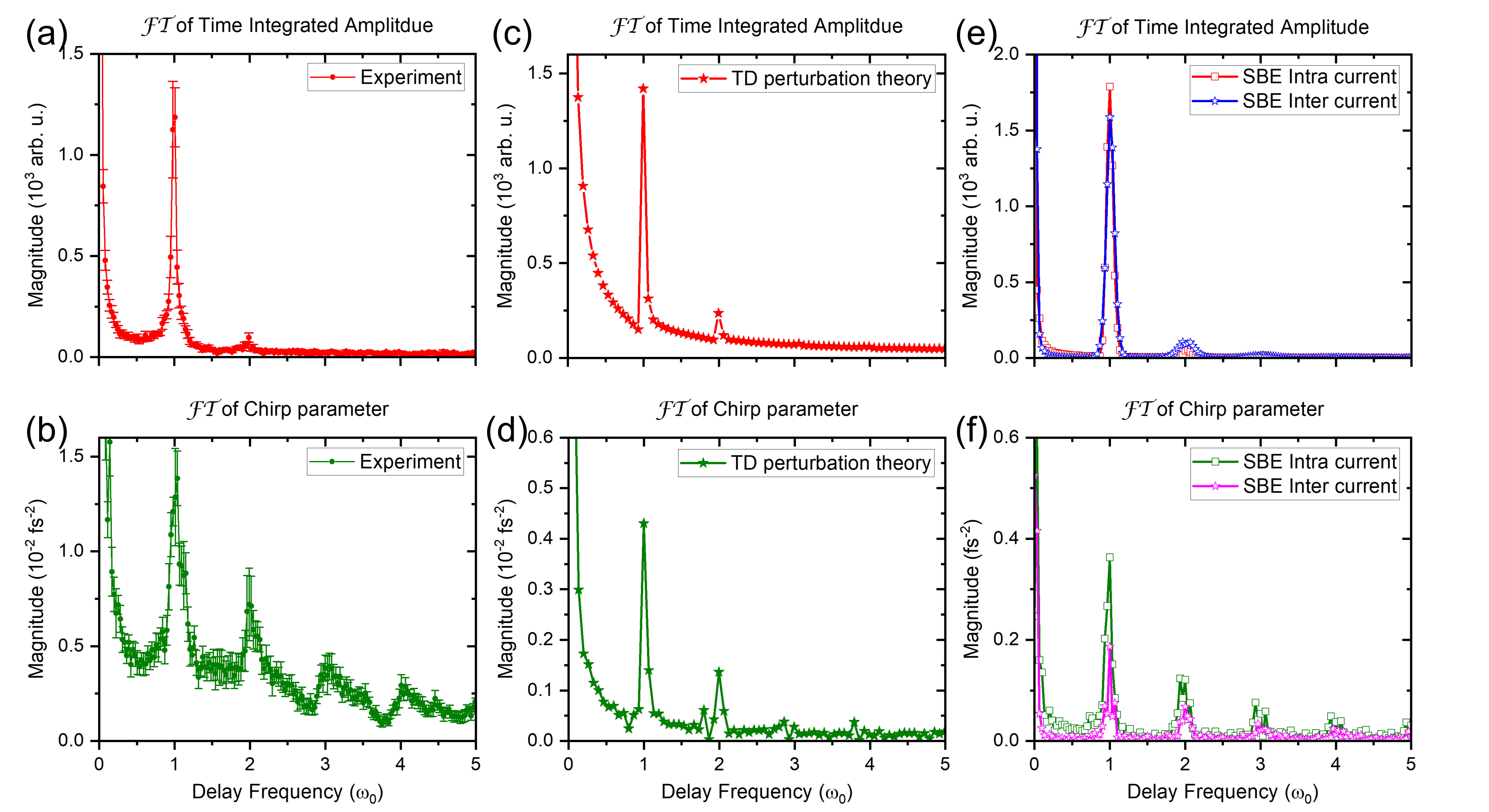}
    \caption{\textbf{Fourier Transformation in Time Delay}: The Fourier transforms ($\mathcal{FT}$) with respect to the time delay of the experimentally measured (a) time-integrated amplitude and (b) temporal chirp parameter are shown. These results are compared with the corresponding (c) time-integrated amplitude and (d) chirp parameter obtained from the electric field calculated using time-dependent (TD) perturbation theory, which includes the band-structure modulation induced by the strong field (see the main text for details). The calculation parameters are: IR pulse intensity $I_0 = 0.2\ \mathrm{TW/cm^2}$, photon energy $\hbar\omega_0 = 1.56\ \mathrm{eV}$, and MgO bandgap $\Delta = 7.78\ \mathrm{eV}$. The (e) time-integrated amplitude and (f) temporal chirp are also extracted from the calculated output electric field using the Semiconductor Bloch Equations (SBE), where the contributions from intraband and interband currents are shown separately.}
    \label{fig: FFT_intensity_chirp}
\end{figure}

We perform a Fourier transformation ($\mathcal{FT}$) of our field observables in the time-delay domain. Figure~\ref{fig: FFT_intensity_chirp} (a) and (b) show the $\mathcal{FT}$ of the time-integrated amplitude and the chirp of the measured DFWM electric field shown in Figure~\ref{fig: exp_intensity_chirp}, respectively. We emphasize that the Fourier transformation is in the time-delay $\tau$ domain and not in real time $t$, and the observed frequency components reflect delay-dependent modulations. We compare these with the theoretically calculated field observables after the inclusion of ultrafast modulation of the band structure within time-dependent perturbation theory in Fig.~\ref{fig: FFT_intensity_chirp} (c) and (d). The calculation shows good agreement with the experimentally measured field observables for an input intensity of 0.2 TW/cm$^2$, which is close to the experimentally measured intensity of $\sim$ 0.1 TW/cm$^2$. The uncertainty in experimentally measured intensity is typically large for femtosecond pulses due to a complex beam profile at the focal spot. The time-integrated amplitude exhibits a strong $1\omega_0$ component of the delay frequency ($\omega_0$ is the central frequency of the NIR pulse). This is because the amplitude of the net incoming electric field modulates with the time delay between the Gate and the Probe pulses. As a result, the nonlinear response function is modulated via adiabatic field-dressed bands (see supplementary material section A for details). The chirp, as shown in Figure~\ref{fig: FFT_intensity_chirp} (b) and (d), is more sensitive to such modulation of the nonlinear response function. Unlike the time-integrated amplitude, the $\mathcal{FT}$ of the chirp shows a significant contribution from higher frequency components ($>1\omega_0$). 


To compare our results to a non-perturbative simulation of our experiment, we solve the semiconductor Bloch equations (SBE) for MgO with our experimental parameters as inputs \cite{Schubert2014, Hohenleutner2015, Huttner2017}. Unlike perturbative models, the non-perturbative SBE framework can separate contributions from intra-band and inter-band currents. The time-delay Fourier transforms of the time-integrated amplitude and the temporal chirp, corresponding to the electric fields generated by intra- and inter-band currents, are shown in Fig.~\ref{fig: FFT_intensity_chirp} (e) and (f). These plots show that the interband and intraband contributions to the DFWM signal-field observables are comparable under our experimental conditions. Further details of the calculation are provided in the supplementary material.

\begin{figure}[t]
    \centering
    \includegraphics[width=\linewidth]{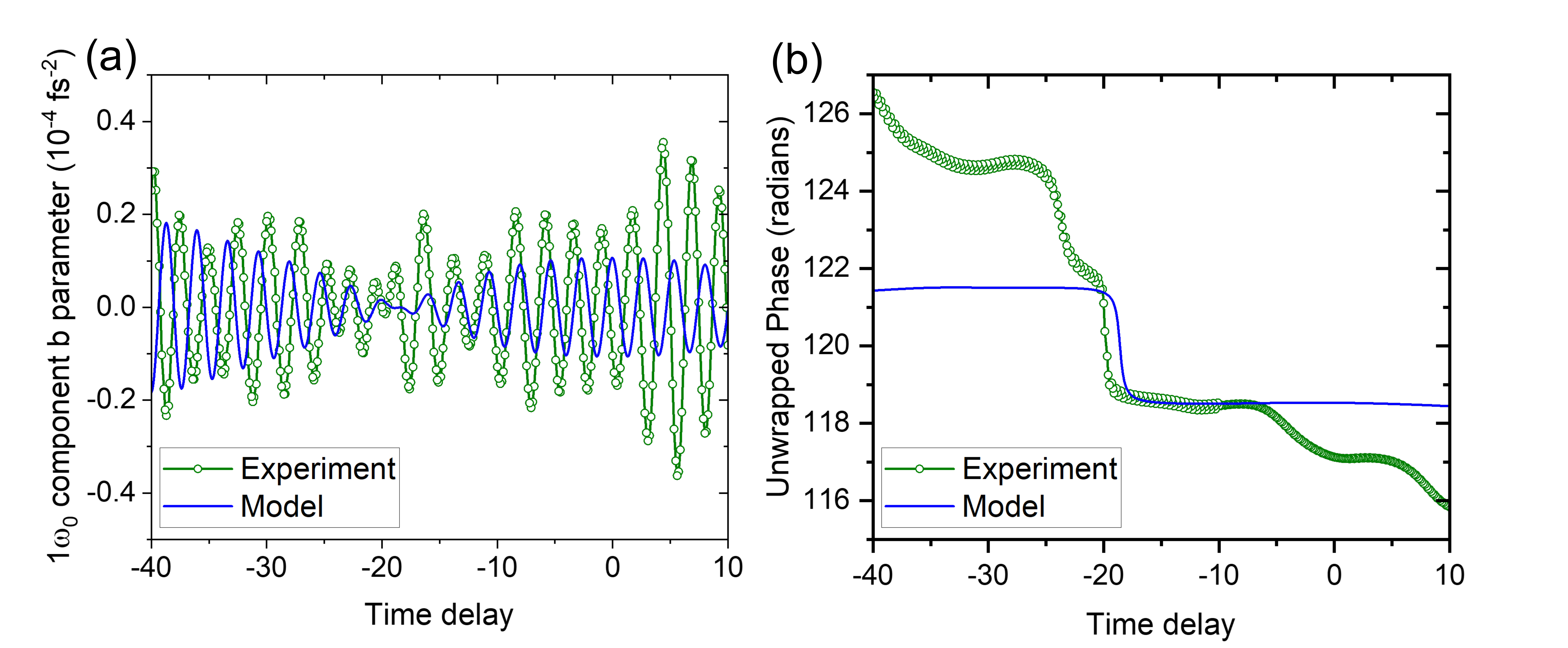}
    \caption{The 1$\omega_0$ component extracted from the Fourier transformation of the temporal chip is isolated, and the inverse Fourier transformation is performed. (a) The oscillation of the 1$\omega_0$ component of the chirp parameter extracted from the experimentally measured DFWM electric field (green) is compared with the nonlinear susceptibility modulation model (magenta) as a function of time delay. (b) The time-dependent unwrapped phases of the oscillations in (a) are also presented, showing that the susceptibility modulation model accurately reproduces the $\pi$-phase jump near $\tau = -20~\mathrm{fs}$, consistent with the experimental observations.}
    \label{fig: Chirp_1w0_amp_phaseq}
\end{figure}

The simple model of nonlinear susceptibility modulation discussed earlier provides physical insight into the emergence of peaks at integer multiples of $\omega_0$ observed in both the experimentally measured and theoretically calculated field observables in Fig.~\ref{fig: FFT_intensity_chirp}. We start from Eq.~\ref{P_higher_order}, where the third-order susceptibility ($\chi^{(3)}(\mathcal{E})$)  is written as an instantaneous function of the total interacting electric field $\mathcal{E}$. The total electric field $\mathcal{E}$ is the sum of the two gate pulses ($E_1$, $E_2$) and the probe pulse ($E_3$) electric field. With an introduction of time delay ($\tau$) between $E_1$ (phase-locked with $E_2$) and $E_3$, the three pulses $E_1, E_2$ and $E_3$ in the frequency domain can be written as,
\begin{subequations}
\begin{align}
    E_1(t-\tau) &\leftrightarrow E_1(\omega)e^{i(\bm{k}_1\cdot \bm{r} - \omega\tau)} + c.c\\
    E_2(t-\tau) &\leftrightarrow E_2(\omega)e^{i(\bm{k}_2\cdot \bm{r} - \omega\tau)} + c.c\\
    E_3(t) &\leftrightarrow E_3(\omega)e^{i\bm{k}_3\cdot \bm{r}}+ c.c.
\end{align}
\end{subequations}
where $\bm{k}_i$ is the wave vector of the $i^{th}$ field. We also expand the modified nonlinear susceptibility in the power series of $\mathcal{E}$ (net electric field) as $\chi^{(3)}(\mathcal{E})=\chi^{(3)}_0+a\mathcal{E}+b\mathcal{E}^2+\mathcal{O}(\mathcal{E}^3)$, where $\chi^{(3)}_0$ is the field-free third order susceptibility. The parameters $a$ and $b$ represent the coefficients governing the first- and second-order modulations of the nonlinear susceptibility. Plugging in all the relations, we get the expression of an effective DFWM polarization as,
\begin{align}
\label{sus_modulation}
    P^{(3)}_{eff} = &\left[\chi^{(3)}_0 + a\left(E_1(\omega)e^{-i\omega\tau} + E_2(\omega)e^{-i\omega\tau} + E_3(\omega)+c.c.\right) \right.\notag\\
    &+\left. b\left(E_1(\omega)e^{-i\omega\tau} + E_2(\omega)e^{-i\omega\tau} + E_3(\omega)+c.c.\right)^2 + \cdots \right]E_1(\omega)E_2^*(\omega)E_3(\omega) e^{i(\bm{k}_1-\bm{k}_2+\bm{k}_3)\cdot \bm{r}}.
\end{align}
Here, it is important to note that the phase-matching conditions arising from terms that include the wave vectors $\bm{k_i}$ corresponding to the input fields only apply to the wave-mixing terms outside the brackets. We do not include wave-vector terms in the field-dependent expansion of $\chi^{(3)}$ because the field-driven modification of $\chi^{(3)}$ is an instantaneous field effect independent of the direction of propagation of the field(s). This allows us to model non-perturbative field modification of the response function within a third-order perturbative model and is distinct from a higher-order perturbation expansion shown in Eq. \ref{no_mod_pol}. Using this expression, we can also derive an analytical expression for the temporal chirp (see the supplementary material for a detailed derivation).

\begin{figure}[ht]
    \centering
    \includegraphics[width=1.0\linewidth]{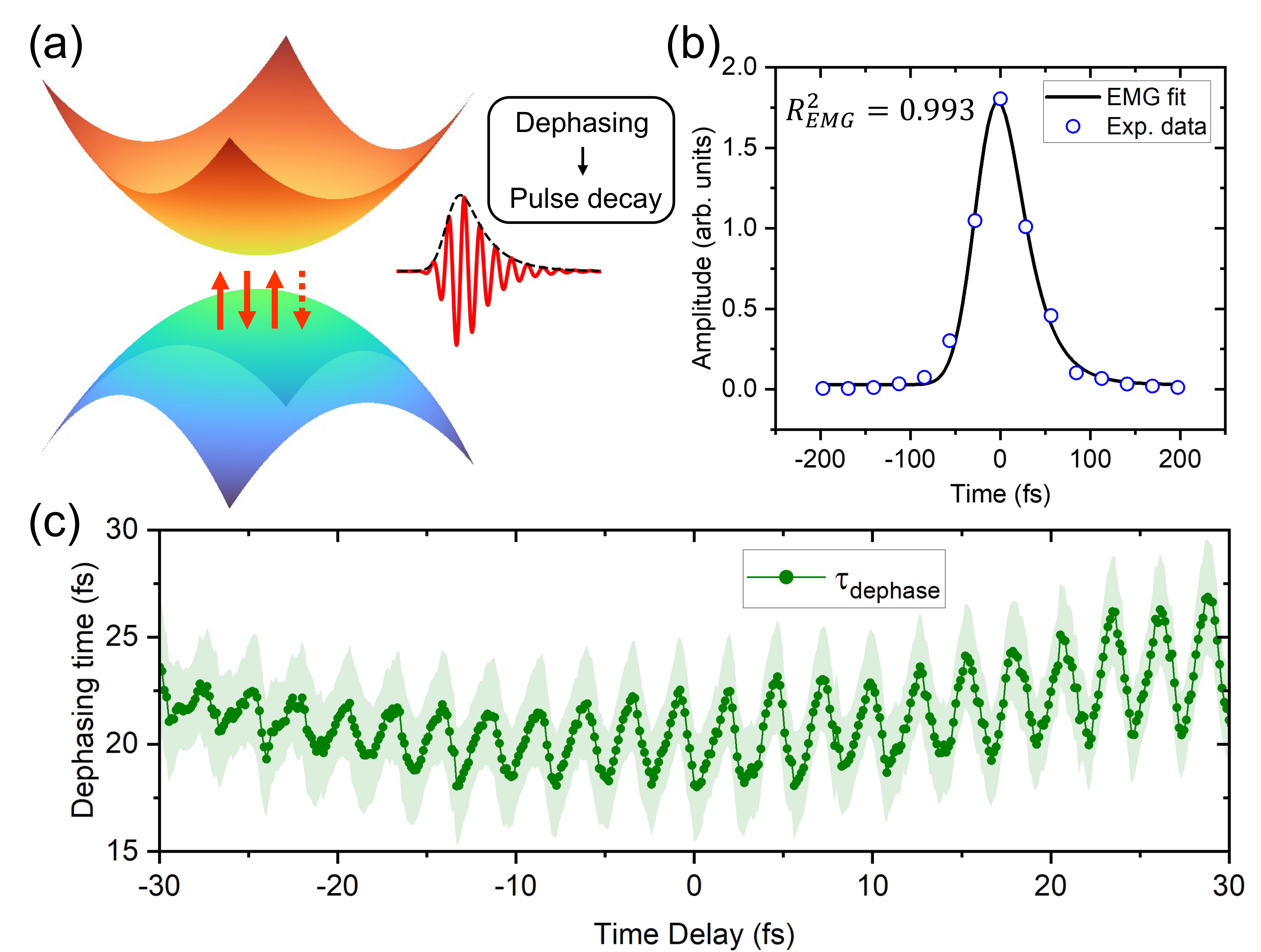}
    \caption{\textbf{Extraction of dephasing times}: (a) Shows a schematic of DFWM in MgO, where the asymmetry in the output pulse contains the information of third-order coherence dephasing. The asymmetry shown in the output pulse is exaggerated for illustrative purposes. (b) A representative plot of the real-time amplitude of the measured output electric field at $\tau = 26$ fs fitted with an exponentially modified Gaussian (EMG) function (black, solid curve) is shown. See supplementary material for details. (c) Dephasing times as a function of time delay are extracted, showing that strong light-matter interaction leads to attosecond modulation and control of dephasing. The shaded region represents the fitting error.}
    \label{fig: fitted_params}
\end{figure}

It can be readily seen from Eq. \ref{sus_modulation} that the $1\omega_0$ peak in Fig.~\ref{fig: FFT_intensity_chirp} primarily originates from the linear expansion of the net field-dependent $\chi^{(3)}$. The second-order expansion contributes not only to the $1\omega_0$ peak but also generates a component at $2\omega_0$. A similar reasoning extends naturally to the higher-order expansions of $\chi^{(3)}$. Within this framework, the difference in the relative peak strengths at $1\omega_0$ and $2\omega_0$ can be understood by noting that the third-order susceptibility is predominantly governed by the linear term, i.e., $\chi^{(3)} \approx \chi_0^{(3)} + a\mathcal{E}$. Analogous phenomena have previously been observed in which an external field modifies both the linear and nonlinear response functions of a material. A well-known example is the optical Kerr effect, in which the change in the linear refractive index of a medium is proportional to the intensity of the applied field \cite{stolen1973, Schimpf2009, bree2011}. Similar effects have also been utilized in electric-field-induced second-harmonic generation, where the external field breaks the inversion symmetry of otherwise centrosymmetric systems, making the second-order response function nonzero\cite{Widhalm2022, Yanagimoto2025}. However, in this experiment, the observed oscillation cannot be explained by including a fifth-order term under the phase-matching condition, as shown previously using Eq.~\ref{no_mod_pol}. To further support this interpretation, we isolate the $1\omega_0$ peak of the measured chirp in Fig.~\ref{fig: FFT_intensity_chirp} (b) and inverse Fourier transform to obtain the single frequency filtered oscillation of the temporal chirp, shown in Fig.~\ref{fig: Chirp_1w0_amp_phaseq} (a). This is then compared with the $1\omega_0$ component of the chirp calculated using Eq.~\ref{sus_modulation}. Fig.~\ref{fig: Chirp_1w0_amp_phaseq} (b) shows the time delay-dependent phase of the oscillations in Fig.~\ref{fig: Chirp_1w0_amp_phaseq} (a). The model accurately reproduces the $\pi$-phase jump in the chirp oscillation near $\tau = 20~\mathrm{fs}$, in excellent agreement with the experimental observation, as shown in Fig.~\ref{fig: Chirp_1w0_amp_phaseq} (b). The offset of this phase jump from zero time delay is due to the chirp present in the input fields.

\subsection{Sub-cycle control of dephasing}

From perturbation theory, the third-order polarization can be written as\cite{Boyd2008, mukamel1995principles},
\begin{align}
     P^{(3)}(t) &= \operatorname{Tr}\left[\hat{\mu}\hat{\rho}^{(3)}(t)e^{-t/\tau_{\text{dephase}}}\right]
\end{align}
In the density matrix formalism, the diagonal terms correspond to the population of electronic states, and the off-diagonal elements represent coherence between states. As coherences evolve in time phase fluctuations attenuate their temporal evolution, a process called dephasing \cite{mukamel1995principles}. To model dephasing of coherences, phenomenological dephasing terms $\gamma_{\text{dephase}} = 1/\tau_{\text{dephase}}$ are added. This means that the output-field pulse width depends on the dephasing of the coherence created by the probe pulse in the final step of the nonlinear interaction. To quantify third-order coherence dephasing, we fit the measured output temporal field with an Exponentially Modified Gaussian (EMG) function. We fix the Gaussian width ($\sigma$) to 21.2 fs (corresponding to the input probe pulse FWHM of 50 fs) and extract the dephasing time $\tau$ of the DFWM signal field as shown in Fig.~\ref{fig: fitted_params} (b) (see supplementary materials for details on fitting). The extracted dephasing times $\tau_{\text{dephase}}$ from the EMG fit are plotted as a function of time delay in Fig.~\ref{fig: fitted_params} (c). The data indicate that band-structure modification resulting from strong light-matter interaction modulates the dephasing times of nonlinear interactions on sub-cycle time scales. Multiple factors could influence dephasing times in solids, including doping, electron-electron interactions, and electron-phonon interactions. In this study, the measured dephasing time likely reflects electron-electron interactions that affect real-space coherence of electron-hole pairs across lattice sites, as in the case of HHG \cite{Brown2024}. Furthermore, band-structure modulation induced by strong light-matter interaction also alters electron-electron interactions, thereby modulating dephasing times.


\section{Conclusion}\label{sec6}

In summary, we have measured the complete electric field of the degenerate four-wave mixing (DFWM) signal in magnesium oxide using spectral interferometry. By analyzing field observables, such as the time-integrated amplitude and the signal field chirp, as functions of the sub-cycle time delay between incoming pulses, we identified unique oscillatory features. We attribute these features to the ultrafast modulation of the nonlinear response function of the crystal due to modulation of the band structure. To support this interpretation, we simulated the DFWM experiment using time-dependent perturbation theory and found that oscillatory behavior could be reproduced only when instantaneous ultrafast band-structure modulation driven by the strong field is included. Such ultrafast band modulation effects are known to leave signatures in other optical observables, such as high-harmonic spectra \cite{uzan2020, uzan2022, Koll2025}. Here, we demonstrate that the influence of field-dressed bands is clearly manifested in the field observables of nonlinear optical signals. Our results indicate that field observables can serve as direct probes of strong-field-driven attosecond electron dynamics in solid-state systems. We have demonstrated that modeling of our experimental observations requires incorporating strong light–matter interactions into the non-perturbative modification of the third-order nonlinear susceptibility. Furthermore, by fitting the real-time electric field envelope of the signal pulses, we have extracted dephasing times corresponding to this non-resonant nonlinear interaction. This represents a direct measurement of dephasing in a nonlinear process via electric field measurement opening doors to enhacing our understanding of dephasing mechanisms. The modulation of the band structure is seen to modulate the dephasing time, resulting in sub-cycle control of dephasing. 

Our findings have significant implications for the manipulation and control of observables governed by the nonlinear response function. For instance, it is well established that signals generated through nonlinear optical processes can serve as sources of non-classical light. Processes such as parametric down-conversion (PDC) and degenerate four-wave mixing (DFWM) are known to produce squeezed light \cite{kumar1984, Breitenbach1997, Adamyan2015, Onodera2022, Fabre2020}. The degree of squeezing in these processes is directly proportional to the nonlinear coupling coefficients of the respective systems, i.e., $\propto \chi^{(2)}$ for PDC and $\propto \chi^{(3)}$ for DFWM \cite{Fabre2020}. It has been demonstrated that squeezing in DFWM signals can also be observed from ultrafast light–matter interactions \cite{Riek2017, Sennary2025, Zimmerman2025}. With the strong sub-cycle modulation of $\chi^{(3)}$ demonstrated in this work, it is now possible to control the degree of squeezing on attosecond time scales \cite{Zimmerman2025}. This could pave the path to achieve extreme levels of squeezing based on field induced nonlinearities that exceed current squeezing levels achieved using periodically-poled nonlinear materials \cite{Vahlbruch2016} with numerous applications in quantum metrology, quantum spectroscopy, and quantum information science.


\section{Methods}

\subsection{Experimental Setup}
\label{exp_methods}

In our setup of the DFWM experiment, as seen in Fig.~\ref{fig1:Expt}, three pulses with different propagation vectors interact with the medium in a non-collinear geometry and generate a fourth pulse due to a third-order nonlinear interaction in a BOXCAR configuration. Due to the non-collinear geometry, the signal pulse is spatially separated from the other three beams. The output signal has a momentum of $\bm{k_s} = \bm{k}_1-\bm{k}_2+\bm{k}_3$. The Probe and Gate beams are centered at 800 nm, with a pulse width of 50 fs and a repetition rate of 1 kHz. The Gate pulse is divided into two phase-locked Gate pulses (Gate 1 and Gate 2) using a mask. The Probe pulse is delayed with respect to the Gate pulses using an optical delay stage, and the Gate polarization can be rotated relative to the Probe polarization. However, in this experiment, the polarizations of all four pulses are fixed along the same direction. The MgO crystal is mounted on a controllable rotation mount to allow precise control of the crystal angle relative to the probe polarization. The Probe and the two Gate pulses each have a pulse energy of $\sim$ 500 nJ and form the input pulses in our DFWM measurement. Before the interaction, a Reference pulse is derived from the Probe pulse. After the interaction, the Signal pulse is isolated by an iris, so the residual Probe and Gate beams are blocked. The Signal pulse passes through a high-contrast polarizer, which extracts the polarization component of the signal along the input polarization direction. The output signal pulse was completely measured using TADPOLE \cite{fittinghoff1996}, where the Reference pulse is combined with the Signal pulse in a spectrometer to obtain the measurement of spectral interference. Frequency-Resolved Optical Gating (FROG) \cite{trebino1993} is used to completely characterize the Reference pulse. The measured spectral interference of the DFWM signal and the Reference can be written as: 
\begin{equation}\label{eq:1}
\begin{split}
S(\omega) &= S_{R}(\omega) + S_{S}(\omega) +\sqrt{S_{R}(\omega)}\sqrt{S_{S}(\omega)}\cos(\varphi_{S}(\omega)-\varphi_{R}(\omega)+\omega\, \tau_{R}) 
\end{split}
\end{equation} 

$S(\omega)$ is the spectral intensity, $\omega$ is the angular frequency, and $\varphi(\omega)$ is the spectral phase. The subscripts $S$ and $R$ stand for signal and reference, respectively. $\tau_{R}$ is the time delay between the Reference and the Signal and the Reference and the background.

\subsection{Simulation}
\label{Theory}
To simulate the complete electric field of the DFWM signal, we employ time-dependent perturbation theory to evaluate the third-order expectation value of the microscopic current operator. The streaking of the electron momentum, $\bm{p}(t) = \bm{p}0 + e\bm{A}_{ext}(t)$, is explicitly incorporated. The emitted DFWM signal is expressed as $\bm{E}_{\mathrm{DFWM}}(t) \propto \frac{\partial}{\partial t}\langle \bm{J}(t) \rangle^{(3)}$, where
\begin{eqnarray}
    \langle J^i (t)\rangle^{(3)} = &&\sum_{\alpha, \beta, \gamma, \bm{q}\in \mathrm{BZ}}\mathcal{N}^{i\alpha\beta\gamma}_{\mathbf{q}} \int_{t_0}^{t} dt_3 \int_{t_0}^{t_3} dt_2 \int_{t_0}^{t_2} dt_1\ A_{ext}^{\alpha}(t_1) A_{ext}^{\beta}(t_2) A_{ext}^{\gamma}(t_3) \notag\\
        &&\times \exp\left(i\int_{t_0}^{t}\Theta_{0,1,\mathbf{q}}(t') dt' + i\int_{t_0}^{t_1}\Theta_{1,0,\mathbf{q}}(t')dt' \right. + \left. i \int_{t_0}^{t_2}\Theta_{0,1,\mathbf{q}}(t')dt' + i\int_{t_0}^{t_3}\Theta_{1,0,\mathbf{q}}(t')dt' \right).
\end{eqnarray}
Here $\Theta_{1,0,\mathbf{q}}(t)$ is the time-dependent energy gap, defined as $\Theta_{1,0,\mathbf{q}}(t) = \epsilon_{\mathrm{CB}}(t) - \epsilon_{\mathrm{VB}}(t)$. Here, $0$ and $1$ denote the valence band (VB) and conduction band (CB), respectively, in the two-band model, and $i, \alpha, \beta, \gamma$ index the polarization directions. Further details of the derivation are provided in the Supplementary Material.

\section*{Acknowledgments}

This work was primarily supported by the U.S. Department of Energy, Office of Science, Basic Energy Sciences, under Award $\#$ DE-SC0024234. NS acknowledges support from the National Science Foundation under award $\#$ 2208061, which supported YZ. Work at the Molecular Foundry was supported by the Office of Science, Office of Basic Energy Sciences, of the U.S. Department of Energy under Contract No. DE-AC02-05CH11231. This research used resources of the National Energy Research Scientific Computing Center (NERSC), a Department of Energy User Facility (project m4269). The authors thank Dr. Rudro Biswas for helpful discussions. 

\bibliography{ref}

@book{Boyd2008,
   author = {Robert W Boyd},
   city = {USA},
   edition = {3rd},
   isbn = {0123694701},
   publisher = {Academic Press, Inc.},
   title = {Nonlinear Optics, Third Edition},
   year = {2008},
}

@article{PhysRevB.57.8774,
  title = {Biexcitonic four-wave-mixing signal in quantum wells: Photon-echo versus free-induction decay},
  author = {Nickolaus, H. and Henneberger, F.},
  journal = {Phys. Rev. B},
  volume = {57},
  issue = {15},
  pages = {8774--8777},
  numpages = {0},
  year = {1998},
  month = {Apr},
  publisher = {American Physical Society},
  doi = {10.1103/PhysRevB.57.8774},
  url = {https://link.aps.org/doi/10.1103/PhysRevB.57.8774}
}

@article{10.1063/1.93802,
    author = {Miller, D. A. B. and Chemla, D. S. and Eilenberger, D. J. and Smith, P. W. and Gossard, A. C. and Wiegmann, W.},
    title = "{Degenerate four‐wave mixing in room‐temperature GaAs/GaAlAs multiple quantum well structures}",
    journal = {Applied Physics Letters},
    volume = {42},
    number = {11},
    pages = {925-927},
    year = {1983},
    month = {06},
    issn = {0003-6951},
    doi = {10.1063/1.93802},
    url = {https://doi.org/10.1063/1.93802},
}

@article{PhysRevB.67.035329,
  title = {Intraband polarization as the source of degenerate four-wave mixing signals in asymmetric semiconductor quantum well structures},
  author = {Dignam, Marc and Hawton, M.},
  journal = {Phys. Rev. B},
  volume = {67},
  issue = {3},
  pages = {035329},
  numpages = {17},
  year = {2003},
  month = {Jan},
  publisher = {American Physical Society},
  doi = {10.1103/PhysRevB.67.035329},
  url = {https://link.aps.org/doi/10.1103/PhysRevB.67.035329}
}

@article{Flayyih2013,
   author = {Ahmed H Flayyih and Amin H Al-Khursan},
   doi = {10.1088/0022-3727/46/44/445102},
   issn = {0022-3727},
   issue = {44},
   journal = {Journal of Physics D: Applied Physics},
   month = {11},
   pages = {445102},
   title = {Theory of four-wave mixing in quantum dot semiconductor optical amplifiers},
   volume = {46},
   year = {2013},
}

@article{10.1063/1.371183,
    author = {Yu, Sungkyu and Lee, Joo In and Viswanath, Annamraju Kasi},
    title = "{Time-resolved four-wave mixing signal in thick bulk GaAs}",
    journal = {Journal of Applied Physics},
    volume = {86},
    number = {6},
    pages = {3159-3164},
    year = {1999},
    month = {09},
    issn = {0021-8979},
    doi = {10.1063/1.371183},
    url = {https://doi.org/10.1063/1.371183},
}

@article{PhysRevB.48.9426,
  title = {Theory of propagation effects in time-resolved four-wave mixing},
  author = {Schillak, P. and Balslev, I.},
  journal = {Phys. Rev. B},
  volume = {48},
  issue = {13},
  pages = {9426--9433},
  numpages = {0},
  year = {1993},
  month = {Oct},
  publisher = {American Physical Society},
  doi = {10.1103/PhysRevB.48.9426},
  url = {https://link.aps.org/doi/10.1103/PhysRevB.48.9426}
}

@article{PhysRevB.51.10601,
  title = {Pulse propagation and many-body effects in semiconductor four-wave mixing},
  author = {Schulze, A. and Knorr, A. and Koch, S. W.},
  journal = {Phys. Rev. B},
  volume = {51},
  issue = {16},
  pages = {10601--10609},
  numpages = {0},
  year = {1995},
  month = {Apr},
  publisher = {American Physical Society},
  doi = {10.1103/PhysRevB.51.10601},
  url = {https://link.aps.org/doi/10.1103/PhysRevB.51.10601}
}

@article{Bencivenga2015,
   author = {F Bencivenga and R Cucini and F Capotondi and A Battistoni and R Mincigrucci and E Giangrisostomi and A Gessini and M Manfredda and I P Nikolov and E Pedersoli and E Principi and C Svetina and P Parisse and F Casolari and M B Danailov and M Kiskinova and C Masciovecchio},
   doi = {10.1038/nature14341},
   issn = {1476-4687},
   issue = {7546},
   journal = {Nature},
   pages = {205-208},
   title = {Four-wave mixing experiments with extreme ultraviolet transient gratings},
   volume = {520},
   url = {https://doi.org/10.1038/nature14341},
   year = {2015},
}

@article{Foglia2018,
  title = {First Evidence of Purely Extreme-Ultraviolet Four-Wave Mixing},
  author = {Foglia, L. and Capotondi, F. and Mincigrucci, R. and Naumenko, D. and Pedersoli, E. and Simoncig, A. and Kurdi, G. and Calvi, A. and Manfredda, M. and Raimondi, L. and Mahne, N. and Zangrando, M. and Masciovecchio, C. and Bencivenga, F.},
  journal = {Phys. Rev. Lett.},
  volume = {120},
  issue = {26},
  pages = {263901},
  numpages = {5},
  year = {2018},
  month = {Jun},
  publisher = {American Physical Society},
  doi = {10.1103/PhysRevLett.120.263901},
  url = {https://link.aps.org/doi/10.1103/PhysRevLett.120.263901}
}

@article{trebino1993,
  title={Using phase retrieval to measure the intensity and phase of ultrashort pulses: frequency-resolved optical gating},
  author={Trebino, Rick and Kane, Daniel J},
  journal={JOSA A},
  volume={10},
  number={5},
  pages={1101--1111},
  year={1993},
  publisher={Optical Society of America}
}

@article{fittinghoff1996,
  title={Measurement of the intensity and phase of ultraweak, ultrashort laser pulses},
  author={Fittinghoff, David N and Bowie, Jason L and Sweetser, John N and Jennings, Richard T and Krumb{\"u}gel, Marco A and DeLong, Kenneth W and Trebino, Rick and Walmsley, Ian A},
  journal={Optics letters},
  volume={21},
  number={12},
  pages={884--886},
  year={1996},
  publisher={Optical Society of America}
}

@article{walz2022,
author = {Francis Walz and Siddhant Pandey and Liang Z. Tan and Niranjan Shivaram},
journal = {Opt. Express},
number = {20},
pages = {36065--36072},
publisher = {Optica Publishing Group},
title = {Electric field measurement of femtosecond time-resolved four-wave mixing signals in molecules},
volume = {30},
month = {Sep},
year = {2022},
url = {https://opg.optica.org/oe/abstract.cfm?URI=oe-30-20-36065},
doi = {10.1364/OE.470925},
}

@book{Bruus2004,
   author = {Henrik Bruus and Karsten Flensberg},
   isbn = {9780198566335},
   publisher = {OUP Oxford},
   title = {Many-Body Quantum Theory in Condensed Matter Physics},
   year = {2004},
}

@article{Vampa2014,
  title = {Theoretical Analysis of High-Harmonic Generation in Solids},
  author = {Vampa, G. and McDonald, C. R. and Orlando, G. and Klug, D. D. and Corkum, P. B. and Brabec, T.},
  journal = {Phys. Rev. Lett.},
  volume = {113},
  issue = {7},
  pages = {073901},
  numpages = {5},
  year = {2014},
  month = {Aug},
  publisher = {American Physical Society},
  doi = {10.1103/PhysRevLett.113.073901},
  url = {https://link.aps.org/doi/10.1103/PhysRevLett.113.073901}
}

@article{Hohenleutner2015,
   author = {M Hohenleutner and F Langer and O Schubert and M Knorr and U Huttner and S W Koch and M Kira and R Huber},
   doi = {10.1038/nature14652},
   issn = {1476-4687},
   issue = {7562},
   journal = {Nature},
   pages = {572-575},
   title = {Real-time observation of interfering crystal electrons in high-harmonic generation},
   volume = {523},
   url = {https://doi.org/10.1038/nature14652},
   year = {2015},
}

@article{Ghimire2019,
   author = {Shambhu Ghimire and David A. Reis},
   doi = {10.1038/s41567-018-0315-5},
   issn = {1745-2473},
   issue = {1},
   journal = {Nature Physics},
   month = {1},
   pages = {10-16},
   title = {High-harmonic generation from solids},
   volume = {15},
   year = {2019},
}

@article{gruzdev2018,
  title = {Ultrafast modification of band structure of wide-band-gap solids by ultrashort pulses of laser-driven electron oscillations},
  author = {Gruzdev, Vitaly and Sergaeva, Olga},
  journal = {Phys. Rev. B},
  volume = {98},
  issue = {11},
  pages = {115202},
  numpages = {15},
  year = {2018},
  month = {Sep},
  publisher = {American Physical Society},
  doi = {10.1103/PhysRevB.98.115202},
  url = {https://link.aps.org/doi/10.1103/PhysRevB.98.115202}
}

@article{keldysh1958influence,
  title={Influence of the lattice vibrations of a crystal on the production of electron-hole pairs in a strong electrical field},
  author={Keldysh, LV},
  journal={Soviet Physics JETP},
  volume={7},
  number={4},
  pages={665--669},
  year={1958}
}

@article{keldysh1958effect,
  title={The effect of a strong electric field on the optical properties of insulating crystals},
  author={Keldysh, LV},
  journal={Sov. Phys. JETP},
  volume={7},
  number={5},
  pages={788--790},
  year={1958}
}

@article{Ghimire2011,
  title = {Observation of high-order harmonic generation in a bulk crystal},
  volume = {7},
  ISSN = {1745-2481},
  url = {http://dx.doi.org/10.1038/nphys1847},
  DOI = {10.1038/nphys1847},
  number = {2},
  journal = {Nature Physics},
  publisher = {Springer Science and Business Media LLC},
  author = {Ghimire,  Shambhu and DiChiara,  Anthony D. and Sistrunk,  Emily and Agostini,  Pierre and DiMauro,  Louis F. and Reis,  David A.},
  year = {2011},
  month = dec,
  pages = {138–141}
}

@article{Mitra2024,
   author = {Sambit Mitra and Álvaro Jiménez-Galán and Mario Aulich and Marcel Neuhaus and Rui E. F. Silva and Volodymyr Pervak and Matthias F. Kling and Shubhadeep Biswas},
   doi = {10.1038/s41586-024-07244-z},
   issn = {0028-0836},
   issue = {8009},
   journal = {Nature},
   month = {4},
   pages = {752-757},
   title = {Light-wave-controlled Haldane model in monolayer hexagonal boron nitride},
   volume = {628},
   year = {2024}
}

@article{
lucchini2016,
author = {M. Lucchini  and S. A. Sato  and A. Ludwig  and J. Herrmann  and M. Volkov  and L. Kasmi  and Y. Shinohara  and K. Yabana  and L. Gallmann  and U. Keller },
title = {Attosecond dynamical Franz-Keldysh effect in polycrystalline diamond},
journal = {Science},
volume = {353},
number = {6302},
pages = {916-919},
year = {2016},
doi = {10.1126/science.aag1268},
URL = {https://www.science.org/doi/abs/10.1126/science.aag1268},
eprint = {https://www.science.org/doi/pdf/10.1126/science.aag1268}}

@article{Wu2015,
   author = {Mengxi Wu and Shambhu Ghimire and David A. Reis and Kenneth J. Schafer and Mette B. Gaarde},
   doi = {10.1103/PhysRevA.91.043839},
   issn = {1050-2947},
   issue = {4},
   journal = {Physical Review A},
   month = {4},
   pages = {043839},
   title = {High-harmonic generation from Bloch electrons in solids},
   volume = {91},
   year = {2015}
}

@article{PhysRevX.7.021017,
  title = {Wannier-Bloch Approach to Localization in High-Harmonics Generation in Solids},
  author = {Osika, Edyta N. and Chac\'on, Alexis and Ortmann, Lisa and Su\'arez, Noslen and P\'erez-Hern\'andez, Jose Antonio and Szafran, Bart\l{}omiej and Ciappina, Marcelo F. and Sols, Fernando and Landsman, Alexandra S. and Lewenstein, Maciej},
  journal = {Phys. Rev. X},
  volume = {7},
  issue = {2},
  pages = {021017},
  numpages = {14},
  year = {2017},
  month = {May},
  publisher = {American Physical Society},
  doi = {10.1103/PhysRevX.7.021017},
  url = {https://link.aps.org/doi/10.1103/PhysRevX.7.021017}
}

@article{PhysRevB.100.195201,
  title = {High harmonic generation in crystals using maximally localized Wannier functions},
  author = {Silva, R. E. F. and Mart\'{\i}n, F. and Ivanov, M.},
  journal = {Phys. Rev. B},
  volume = {100},
  issue = {19},
  pages = {195201},
  numpages = {6},
  year = {2019},
  month = {Nov},
  publisher = {American Physical Society},
  doi = {10.1103/PhysRevB.100.195201},
  url = {https://link.aps.org/doi/10.1103/PhysRevB.100.195201}
}

@article{Schubert2014,
   author = {O. Schubert and M. Hohenleutner and F. Langer and B. Urbanek and C. Lange and U. Huttner and D. Golde and T. Meier and M. Kira and S. W. Koch and R. Huber},
   doi = {10.1038/nphoton.2013.349},
   issn = {1749-4885},
   issue = {2},
   journal = {Nature Photonics},
   month = {2},
   pages = {119-123},
   title = {Sub-cycle control of terahertz high-harmonic generation by dynamical Bloch oscillations},
   volume = {8},
   year = {2014}
}

@article{Huttner2017,
   author = {Ulrich Huttner and Mackillo Kira and Stephan W. Koch},
   doi = {10.1002/lpor.201700049},
   issn = {1863-8880},
   issue = {4},
   journal = {Laser \& Photonics Reviews},
   month = {7},
   title = {Ultrahigh Off‐Resonant Field Effects in Semiconductors},
   volume = {11},
   year = {2017}
}

@article{Jauho1996,
  title = {Dynamical Franz-Keldysh Effect},
  author = {Jauho, A. P. and Johnsen, K.},
  journal = {Phys. Rev. Lett.},
  volume = {76},
  issue = {24},
  pages = {4576--4579},
  numpages = {0},
  year = {1996},
  month = {Jun},
  publisher = {American Physical Society},
  doi = {10.1103/PhysRevLett.76.4576},
  url = {https://link.aps.org/doi/10.1103/PhysRevLett.76.4576}
}

@article{Otobe2016,
  title = {Femtosecond time-resolved dynamical Franz-Keldysh effect},
  author = {Otobe, T. and Shinohara, Y. and Sato, S. A. and Yabana, K.},
  journal = {Phys. Rev. B},
  volume = {93},
  issue = {4},
  pages = {045124},
  numpages = {9},
  year = {2016},
  month = {Jan},
  publisher = {American Physical Society},
  doi = {10.1103/PhysRevB.93.045124},
  url = {https://link.aps.org/doi/10.1103/PhysRevB.93.045124}
}

@article{Han2019,
   author = {Seunghwoi Han and Lisa Ortmann and Hyunwoong Kim and Yong Woo Kim and Takashi Oka and Alexis Chacon and Brent Doran and Marcelo Ciappina and Maciej Lewenstein and Seung-Woo Kim and Seungchul Kim and Alexandra S. Landsman},
   doi = {10.1038/s41467-019-11096-x},
   issn = {2041-1723},
   issue = {1},
   journal = {Nature Communications},
   month = {7},
   pages = {3272},
   title = {Extraction of higher-order nonlinear electronic response in solids using high harmonic generation},
   volume = {10},
   year = {2019}
}

@article{Reislohner2023,
  title = {Dynamical Franz-Keldysh Effect in Diamond in the Deep Ultraviolet Probed by Transient Absorption and Dispersion Spectroscopy Using a Miniature Beamline},
  author = {Reisl\"ohner, Jan and Chen, Xiao and Kim, Doyeong and Botti, Silvana and Pfeiffer, Adrian N.},
  journal = {Phys. Rev. Lett.},
  volume = {131},
  issue = {13},
  pages = {136902},
  numpages = {6},
  year = {2023},
  month = {Sep},
  publisher = {American Physical Society},
  doi = {10.1103/PhysRevLett.131.136902},
  url = {https://link.aps.org/doi/10.1103/PhysRevLett.131.136902}
}

@article{Neufeld2022,
  title = {Time- and angle-resolved photoelectron spectroscopy of strong-field light-dressed solids: Prevalence of the adiabatic band picture},
  author = {Neufeld, Ofer and Mao, Wenwen and H\"ubener, Hannes and Tancogne-Dejean, Nicolas and Sato, Shunsuke A. and De Giovannini, Umberto and Rubio, Angel},
  journal = {Phys. Rev. Res.},
  volume = {4},
  issue = {3},
  pages = {033101},
  numpages = {17},
  year = {2022},
  month = {Aug},
  publisher = {American Physical Society},
  doi = {10.1103/PhysRevResearch.4.033101},
  url = {https://link.aps.org/doi/10.1103/PhysRevResearch.4.033101}
}

@article{wang2025,
  title = {Light-Field Dressing of Transient Photoexcited States above the Fermi Energy},
  author = {Wang, Fei and Chen, Wanying and Bao, Changhua and Lin, Tianyun and Zhong, Haoyuan and Zhang, Hongyun and Zhou, Shuyun},
  journal = {Phys. Rev. Lett.},
  volume = {134},
  issue = {14},
  pages = {146401},
  numpages = {6},
  year = {2025},
  month = {Apr},
  publisher = {American Physical Society},
  doi = {10.1103/PhysRevLett.134.146401},
  url = {https://link.aps.org/doi/10.1103/PhysRevLett.134.146401}
}

@article{tiwari2023,
  title = {Floquet theory and computational method for the optical absorption of laser-dressed solids},
  author = {Tiwari, Vishal and Gu, Bing and Franco, Ignacio},
  journal = {Phys. Rev. B},
  volume = {108},
  issue = {6},
  pages = {064308},
  numpages = {16},
  year = {2023},
  month = {Aug},
  publisher = {American Physical Society},
  doi = {10.1103/PhysRevB.108.064308},
  url = {https://link.aps.org/doi/10.1103/PhysRevB.108.064308}
}

@article{uzan2022,
  title={Observation of light-driven band structure via multiband high-harmonic spectroscopy},
  author={Uzan-Narovlansky, Ayelet J and Jim{\'e}nez-Gal{\'a}n, {\'A}lvaro and Orenstein, Gal and Silva, Rui EF and Arusi-Parpar, Talya and Shames, Sergei and Bruner, Barry D and Yan, Binghai and Smirnova, Olga and Ivanov, Misha and others},
  journal={Nature photonics},
  volume={16},
  number={6},
  pages={428--432},
  year={2022},
  publisher={Nature Publishing Group UK London},
  doi = {https://doi.org/10.1038/s41566-022-01010-1}
}

@article{uzan2020,
  title={Attosecond spectral singularities in solid-state high-harmonic generation},
  author={Uzan, Ayelet Julie and Orenstein, Gal and Jim{\'e}nez-Gal{\'a}n, {\'A}lvaro and McDonald, Chris and Silva, Rui EF and Bruner, Barry D and Klimkin, Nikolai D and Blanchet, Valerie and Arusi-Parpar, Talya and Kr{\"u}ger, Michael and others},
  journal={Nature Photonics},
  volume={14},
  number={3},
  pages={183--187},
  year={2020},
  publisher={Nature Publishing Group UK London},
  doi = {https://doi.org/10.1038/s41566-019-0574-4}
}

@article{chin2001,
  title = {Extreme Midinfrared Nonlinear Optics in Semiconductors},
  author = {Chin, Alan H. and Calder\'on, Oscar G. and Kono, Junichiro},
  journal = {Phys. Rev. Lett.},
  volume = {86},
  issue = {15},
  pages = {3292--3295},
  numpages = {0},
  year = {2001},
  month = {Apr},
  publisher = {American Physical Society},
  doi = {10.1103/PhysRevLett.86.3292},
  url = {https://link.aps.org/doi/10.1103/PhysRevLett.86.3292}
}

@article{gorelova2024,
    author = {Popova-Gorelova, Daria and Santra, Robin},
    title = {Microscopic nonlinear optical response: Analysis and calculations with the Floquet–Bloch formalism},
    journal = {Structural Dynamics},
    volume = {11},
    number = {1},
    pages = {014102},
    year = {2024},
    month = {02},
    issn = {2329-7778},
    doi = {10.1063/4.0000220},
    url = {https://doi.org/10.1063/4.0000220}
}

@article{Nordstrom1998,
  title = {Excitonic Dynamical Franz-Keldysh Effect},
  author = {Nordstrom, K. B. and Johnsen, K. and Allen, S. J. and Jauho, A.-P. and Birnir, B. and Kono, J. and Noda, T. and Akiyama, H. and Sakaki, H.},
  journal = {Phys. Rev. Lett.},
  volume = {81},
  issue = {2},
  pages = {457--460},
  numpages = {0},
  year = {1998},
  month = {Jul},
  publisher = {American Physical Society},
  doi = {10.1103/PhysRevLett.81.457},
  url = {https://link.aps.org/doi/10.1103/PhysRevLett.81.457}
}

@article{Levine1989,
  title = {Linear optical response in silicon and germanium including self-energy effects},
  author = {Levine, Zachary H. and Allan, Douglas C.},
  journal = {Phys. Rev. Lett.},
  volume = {63},
  issue = {16},
  pages = {1719--1722},
  numpages = {0},
  year = {1989},
  month = {Oct},
  publisher = {American Physical Society},
  doi = {10.1103/PhysRevLett.63.1719},
  url = {https://link.aps.org/doi/10.1103/PhysRevLett.63.1719}
}

@article{Sham1983,
  title = {Density-Functional Theory of the Energy Gap},
  author = {Sham, L. J. and Schl\"uter, M.},
  journal = {Phys. Rev. Lett.},
  volume = {51},
  issue = {20},
  pages = {1888--1891},
  numpages = {0},
  year = {1983},
  month = {Nov},
  publisher = {American Physical Society},
  doi = {10.1103/PhysRevLett.51.1888},
  url = {https://link.aps.org/doi/10.1103/PhysRevLett.51.1888}
}

@article{Liu2024,
  title = {High-throughput hybrid-functional DFT calculations of bandgaps and formation energies and multifidelity learning with uncertainty quantification},
  author = {Liu, Mohan and Gopakumar, Abhijith and Hegde, Vinay Ishwar and He, Jiangang and Wolverton, Chris},
  journal = {Phys. Rev. Mater.},
  volume = {8},
  issue = {4},
  pages = {043803},
  numpages = {12},
  year = {2024},
  month = {Apr},
  publisher = {American Physical Society},
  doi = {10.1103/PhysRevMaterials.8.043803},
  url = {https://link.aps.org/doi/10.1103/PhysRevMaterials.8.043803}
}

@article{Crowley2016,
  title = {Resolution of the Band Gap Prediction Problem for Materials Design},
  volume = {7},
  ISSN = {1948-7185},
  url = {http://dx.doi.org/10.1021/acs.jpclett.5b02870},
  DOI = {10.1021/acs.jpclett.5b02870},
  number = {7},
  journal = {The Journal of Physical Chemistry Letters},
  publisher = {American Chemical Society (ACS)},
  author = {Crowley,  Jason M. and Tahir-Kheli,  Jamil and Goddard,  William A.},
  year = {2016},
  month = mar,
  pages = {1198–1203}
}

@article{Shirley1980,
author = {John A. Shirley and Robert J. Hall and Alan C. Eckbreth},
journal = {Opt. Lett.},
number = {9},
pages = {380--382},
publisher = {Optica Publishing Group},
title = {Folded BOXCARS for rotational Raman studies},
volume = {5},
month = {Sep},
year = {1980},
url = {https://opg.optica.org/ol/abstract.cfm?URI=ol-5-9-380},
doi = {10.1364/OL.5.000380},
}

@article{stolen1973,
    author = {Stolen, R.H. and Ashkin, A.},
    title = {Optical Kerr effect in glass waveguide},
    journal = {Applied Physics Letters},
    volume = {22},
    number = {6},
    pages = {294-296},
    year = {1973},
    month = {03},
    issn = {0003-6951},
    doi = {10.1063/1.1654644},
    url = {https://doi.org/10.1063/1.1654644},
}

@article{Schimpf2009,
author = {Damian N. Schimpf and Tino Eidam and Enrico Seise and Steffen H\"{a}drich and Jens Limpert and Andreas T\"{u}nnermann},
journal = {Opt. Express},
number = {21},
pages = {18774--18781},
publisher = {Optica Publishing Group},
title = {Circular versus linear polarization in laser-amplifiers with Kerr-nonlinearity},
volume = {17},
month = {Oct},
year = {2009},
url = {https://opg.optica.org/oe/abstract.cfm?URI=oe-17-21-18774},
doi = {10.1364/OE.17.018774},
}

@article{bree2011,
  title = {Saturation of the All-Optical Kerr Effect},
  author = {Br\'ee, Carsten and Demircan, Ayhan and Steinmeyer, G\"unter},
  journal = {Phys. Rev. Lett.},
  volume = {106},
  issue = {18},
  pages = {183902},
  numpages = {4},
  year = {2011},
  month = {May},
  publisher = {American Physical Society},
  doi = {10.1103/PhysRevLett.106.183902},
  url = {https://link.aps.org/doi/10.1103/PhysRevLett.106.183902}
}

@article{Widhalm2022,
    author = {Alex Widhalm and Christian Golla and Nils Weber and Peter Mackwitz and Artur Zrenner and Cedrik Meier},
    journal = {Opt. Express},
    number = {4},
    pages = {4867--4874},
    publisher = {Optica Publishing Group},
    title = {Electric-field-induced second harmonic generation in silicon dioxide},
    volume = {30},
    month = {Feb},
    year = {2022},
    url = {https://opg.optica.org/oe/abstract.cfm?URI=oe-30-4-4867},
    doi = {10.1364/OE.443489},
}

@article{Yanagimoto2025,
  title = {Programmable on-chip nonlinear photonics},
  ISSN = {1476-4687},
  url = {http://dx.doi.org/10.1038/s41586-025-09620-9},
  DOI = {10.1038/s41586-025-09620-9},
  journal = {Nature},
  publisher = {Springer Science and Business Media LLC},
  author = {Yanagimoto,  Ryotatsu and Ash,  Benjamin A. and Sohoni,  Mandar M. and Stein,  Martin M. and Zhao,  Yiqi and Presutti,  Federico and Jankowski,  Marc and Wright,  Logan G. and Onodera,  Tatsuhiro and McMahon,  Peter L.},
  year = {2025},
  month = oct 
}

@article{Breitenbach1997,
  title = {Measurement of the quantum states of squeezed light},
  volume = {387},
  ISSN = {1476-4687},
  url = {http://dx.doi.org/10.1038/387471a0},
  DOI = {10.1038/387471a0},
  number = {6632},
  journal = {Nature},
  publisher = {Springer Science and Business Media LLC},
  author = {Breitenbach,  G. and Schiller,  S. and Mlynek,  J.},
  year = {1997},
  month = may,
  pages = {471–475}
}

@article{Onodera2022,
  title = {Nonlinear quantum behavior of ultrashort-pulse optical parametric oscillators},
  volume = {105},
  ISSN = {2469-9934},
  url = {http://dx.doi.org/10.1103/PhysRevA.105.033508},
  DOI = {10.1103/physreva.105.033508},
  number = {3},
  journal = {Physical Review A},
  publisher = {American Physical Society (APS)},
  author = {Onodera,  Tatsuhiro and Ng,  Edwin and Gustin,  Chris and L\"{o}rch,  Niels and Yamamura,  Atsushi and Hamerly,  Ryan and McMahon,  Peter L. and Marandi,  Alireza and Mabuchi,  Hideo},
  year = {2022},
  month = mar 
}

@article{Adamyan2015,
  title = {Strong squeezing in periodically modulated optical parametric oscillators},
  volume = {92},
  ISSN = {1094-1622},
  url = {http://dx.doi.org/10.1103/PhysRevA.92.053818},
  DOI = {10.1103/physreva.92.053818},
  number = {5},
  journal = {Physical Review A},
  publisher = {American Physical Society (APS)},
  author = {Adamyan,  Hayk H. and Bergou,  János A. and Gevorgyan,  Narine T. and Kryuchkyan,  Gagik Yu.},
  year = {2015},
  month = nov 
}

@article{Fabre2020,
  title = {Modes and states in quantum optics},
  volume = {92},
  ISSN = {1539-0756},
  url = {http://dx.doi.org/10.1103/RevModPhys.92.035005},
  DOI = {10.1103/revmodphys.92.035005},
  number = {3},
  journal = {Reviews of Modern Physics},
  publisher = {American Physical Society (APS)},
  author = {Fabre,  C. and Treps,  N.},
  year = {2020},
  month = sep 
}

@article{kumar1984,
  title = {Degenerate four-wave mixing as a possible source of squeezed-state light},
  author = {Bondurant, Roy S. and Prem Kumar and Shapiro, Jeffrey H. and Maeda, Mari},
  journal = {Phys. Rev. A},
  volume = {30},
  issue = {1},
  pages = {343--353},
  numpages = {0},
  year = {1984},
  month = {Jul},
  publisher = {American Physical Society},
  doi = {10.1103/PhysRevA.30.343},
  url = {https://link.aps.org/doi/10.1103/PhysRevA.30.343}
}

@article{Sennary2025,
  title = {Attosecond quantum uncertainty dynamics and ultrafast squeezed light for quantum communication},
  volume = {14},
  ISSN = {2047-7538},
  url = {http://dx.doi.org/10.1038/s41377-025-02055-x},
  DOI = {10.1038/s41377-025-02055-x},
  number = {1},
  journal = {Light: Science \& amp; Applications},
  publisher = {Springer Science and Business Media LLC},
  author = {Sennary,  Mohamed and Rivera-Dean,  Javier and ElKabbash,  Mohamed and Pervak,  Vladimir and Lewenstein,  Maciej and Hassan,  Mohammed Th.},
  year = {2025},
  month = oct 
}

@article{Koll2025,
author = {Lisa-Marie Koll and Simon Vendelbo Bylling Jensen and Pieter J. van Essen and Brian de Keijzer and Emilia Olsson and Jon Cottom and Tobias Witting and Anton Husakou and Marc J. J. Vrakking and Lars Bojer Madsen and Peter M. Kraus and Peter J\"{u}rgens},
journal = {Optica},
number = {10},
pages = {1606--1614},
publisher = {Optica Publishing Group},
title = {Extreme ultraviolet high-harmonic interferometry of excitation-induced bandgap dynamics in solids},
volume = {12},
month = {Oct},
year = {2025},
url = {https://opg.optica.org/optica/abstract.cfm?URI=optica-12-10-1606},
doi = {10.1364/OPTICA.559022},
}

@article{Vahlbruch2016,
  title = {Detection of 15 dB Squeezed States of Light and their Application for the Absolute Calibration of Photoelectric Quantum Efficiency},
  author = {Vahlbruch, Henning and Mehmet, Moritz and Danzmann, Karsten and Schnabel, Roman},
  journal = {Phys. Rev. Lett.},
  volume = {117},
  issue = {11},
  pages = {110801},
  numpages = {5},
  year = {2016},
  month = {Sep},
  publisher = {American Physical Society},
  doi = {10.1103/PhysRevLett.117.110801},
  url = {https://link.aps.org/doi/10.1103/PhysRevLett.117.110801}
}

@article{Riek2017,
  title = {Subcycle quantum electrodynamics},
  volume = {541},
  ISSN = {1476-4687},
  url = {http://dx.doi.org/10.1038/nature21024},
  DOI = {10.1038/nature21024},
  number = {7637},
  journal = {Nature},
  publisher = {Springer Science and Business Media LLC},
  author = {Riek,  C. and Sulzer,  P. and Seeger,  M. and Moskalenko,  A. S. and Burkard,  G. and Seletskiy,  D. V. and Leitenstorfer,  A.},
  year = {2017},
  month = jan,
  pages = {376–379}
}

@book{mukamel1995principles,
  title={Principles of nonlinear optical spectroscopy},
  author={Mukamel, Shaul},
  year={1995},
  publisher={Oxford University Press},
  address={New York},
  isbn={978-0195092783}
}

@misc{Zimmerman2025,
  doi = {10.48550/ARXIV.2512.17046},
  url = {https://arxiv.org/abs/2512.17046},
  author = {Zimmerman,  Russell and Kumar,  Shashank and Tiwari,  Shiva Kant and Liu,  Eric and Walz,  Francis and Pandey,  Siddhant and Economou,  George J. and Alaeian,  Hadiseh and Liao,  Chen-Ting and Walther,  Valentin and Shivaram,  Niranjan},
  title = {Attosecond Control of Squeezed Light},
  publisher = {arXiv},
  year = {2025},
  copyright = {Creative Commons Attribution 4.0 International}
}

@article{Brown2024,
  title = {Real-space perspective on dephasing in solid-state high harmonic generation},
  author = {Brown, Graham G. and Jim\'enez-Gal\'an, \'Alvaro and Silva, Rui E. F. and Ivanov, Misha},
  journal = {Phys. Rev. Res.},
  volume = {6},
  issue = {4},
  pages = {043005},
  numpages = {6},
  year = {2024},
  month = {Oct},
  publisher = {American Physical Society},
  doi = {10.1103/PhysRevResearch.6.043005},
  url = {https://link.aps.org/doi/10.1103/PhysRevResearch.6.043005}
}

\newpage

\section{Supplementary information}
\subsection{Exponentially modified Gaussian fitting}
\begin{figure}[h]
    \centering
    \includegraphics[width=0.95\linewidth]{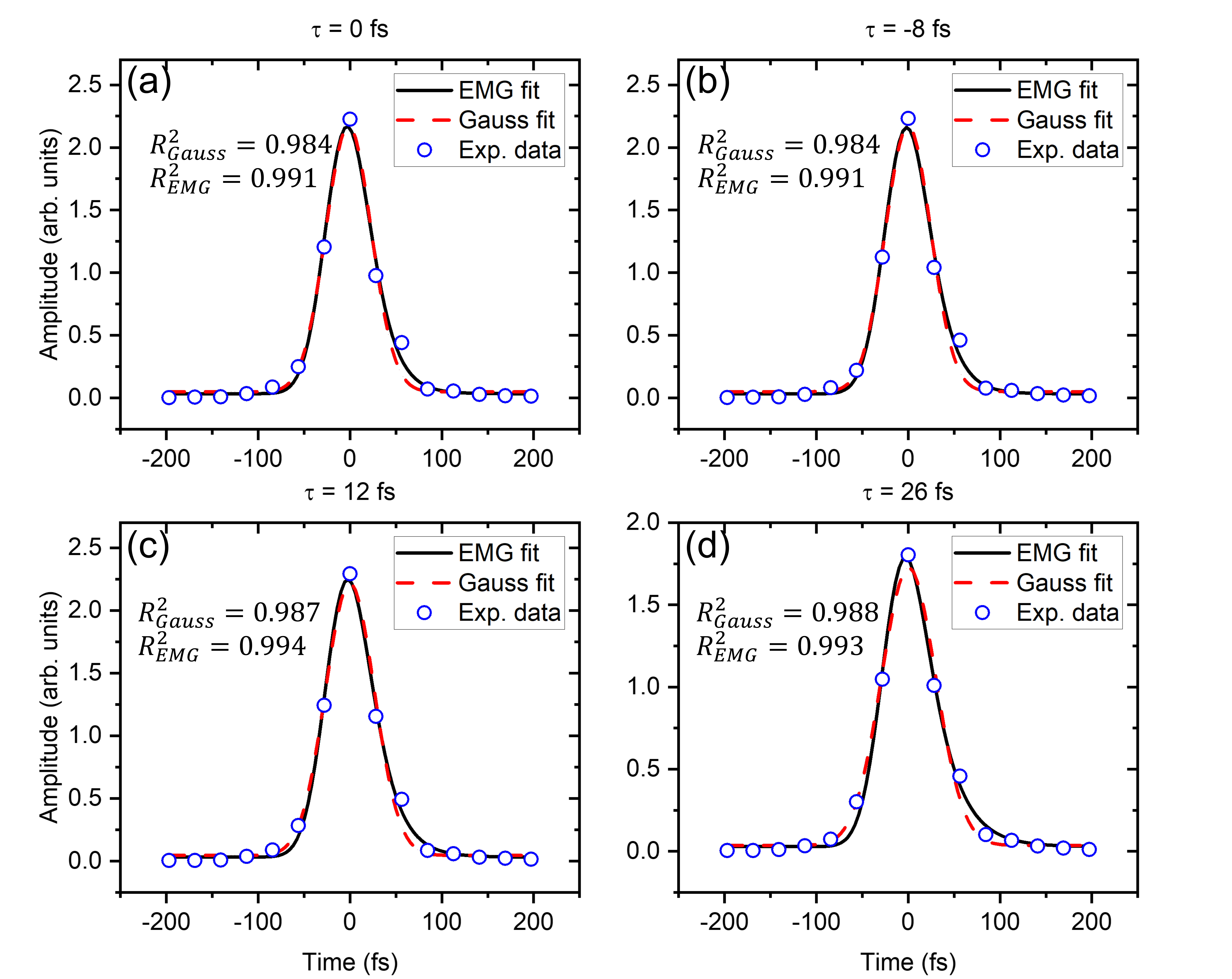}
    \caption{Fitting of temporal amplitude of DFWM signal at (a) $\tau = 0$ fs, (b) $\tau = -8$ fs, (c) $\tau = 12$ fs, and (d) $\tau = 26$ fs using Gaussian (red dashed curve) and Exponentially modified Gaussian (black solid curve) functions. The goodness of both fits are also shown in each plot.}
    \label{fig:fitting}
\end{figure}

\begin{figure}
    \centering
    \includegraphics[width=\linewidth]{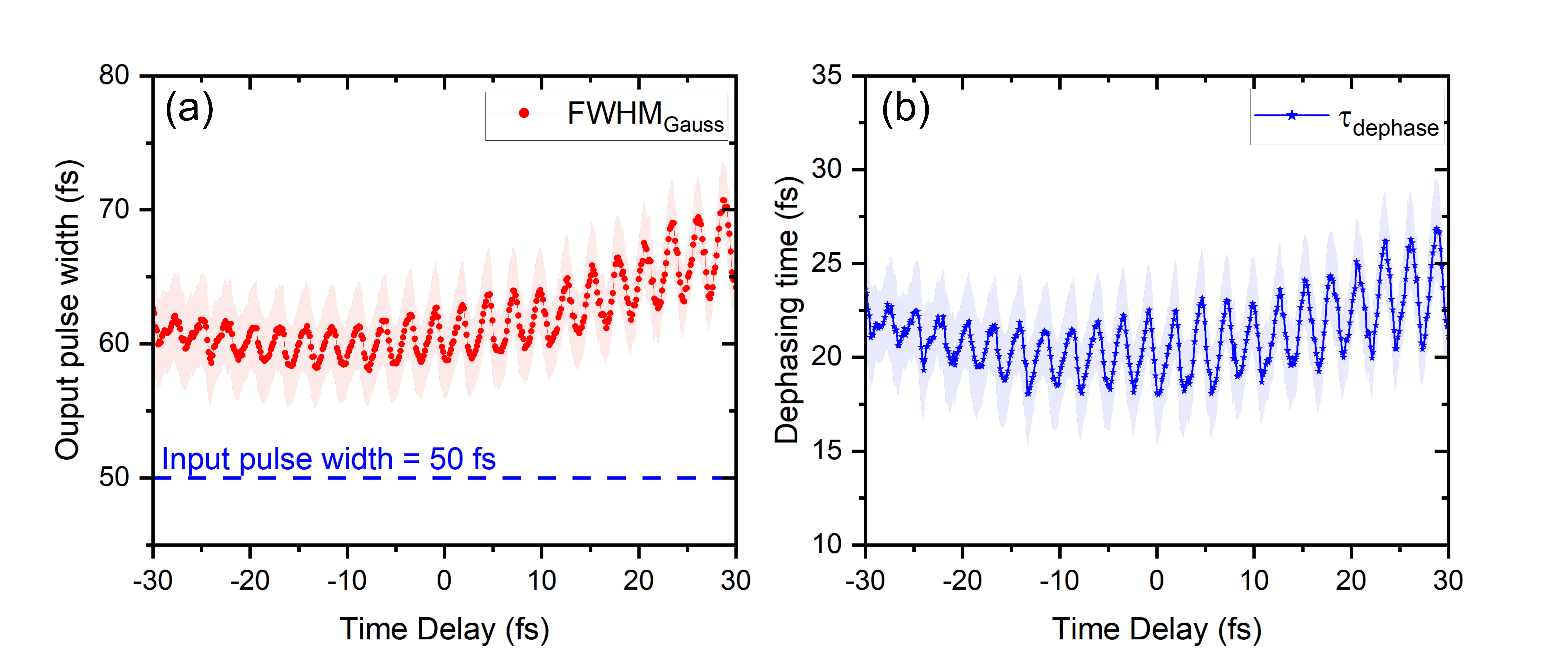}
    \caption{(a) Extracted output pulse width from the Gaussian fit as a function of time delay. (b) Extracted dephasing times $\tau_{\text{dephase}}$ from an Exponentially Modified Gaussian fit with a fixed Gaussian width (FWHM) of 50 fs corresponding to the probe pulse width. The shaded area represents the fitting error.}
    \label{fig:gauss_exp_fitting}
\end{figure}

In Fig.~\ref{fig: exp_intensity_chirp} (a), the DFWM signal amplitude has an approximately Gaussian form in real-time for each time delay. However, a small asymmetry is observed around zero time (see Fig.~\ref{fig:fitting}  due to dephasing of the coherence generated in the final step of the DFWM process. Consequently, for each time delay, we fit both a simple Gaussian profile and an exponentially modified Gaussian (EMG) function. The fitting functions are defined as,
\begin{subequations}\label{fitting_functions}
    \begin{align}
        f_{\text{Gauss}} &= A\exp{\left(-\frac{1}{2}\left(\frac{t-t_0}{\sigma}\right)^2\right)} + C\\
        f_{\text{EMG}} &= A \exp{\left(-\frac{1}{2}\left(\frac{t-t_0}{\sigma}\right)^2\right)} \frac{\sigma}{\tau_{\text{dephase}}} \sqrt{\frac{\pi}{2}} \text{erfcx}\left(\frac{1}{\sqrt{2}}\left(\frac{\sigma}{\tau_{\text{dephase}}}-\frac{t-t_0}{\sigma}\right)\right) + C
    \end{align}
\end{subequations}
where $\sigma$ is the Gaussian width and $\tau_{\text{dephase}}$ corresponds to the dephasing time of the interaction. The scaled complementary error function is defined as $\operatorname{erfcx}(t) = \exp(t^{2})\operatorname{erfc}(t)$, with $\operatorname{erfc}(t)$ being the complementary error function. Fig. \ref{fig:fitting} shows the Gaussian (red dashed curve) and EMG (black solid curves) fitting along with the experimental data (blue markers) with temporal resolution of $28$ fs for a few different time delays. For the EMG fit, we fix the width $\sigma = 21.2$ fs (corresponding to the input probe pulse FWHM of 50 fs). The parameters are extracted from the fits and plotted as a function of time delay in Fig. \ref{fig:gauss_exp_fitting}. The EMG fit shows a higher $R^2$ value and is thus a better model for the signal pulse shape.

\subsection{Nonlinear susceptibility modulation}

The induced polarization spectrum $P(\omega)$ can be expressed as a power series of the electric field as,
\begin{align}
\label{P_chi}
    P(\omega) = \epsilon_0[\chi^{(1)}\mathcal{E}(\omega) + \chi^{(3)}(\omega; \omega_1, \omega_2, \omega_3)\mathcal{E}(\omega_1)\mathcal{E}(\omega_2)\mathcal{E}(\omega_3)+ \cdots]
\end{align}
where $\chi^{(n)}$ is the n$^{th}$ order non-linear susceptibility tensor. The $\mathcal{E}(\omega)$ is the spectrum of the total incoming electric field. Here, we have excluded the even-order terms because they vanish for centrosymmetric materials. For the signal corresponding to a four-wave mixing process, such as DFWM, we are interested in the terms with $n\geq 3$. With an introduction of time delay ($\tau$) between $E_1$ (phase-locked with $E_2$) and $E_3$, the three pulse $E_1, E_2$ and $E_3$ in frequency domain can be written as,
\begin{subequations}
\begin{align}
    E_1(t-\tau) &\leftrightarrow E_1(\omega)e^{i(\bm{k}_1\cdot \bm{r} - \omega\tau)} + c.c\\
    E_2(t-\tau) &\leftrightarrow E_2(\omega)e^{i(\bm{k}_2\cdot \bm{r} - \omega\tau)} + c.c\\
    E_3(t) &\leftrightarrow E_3(\omega)e^{i\bm{k}_3\cdot \bm{r}}+ c.c
\end{align}
\end{subequations}
where $\bm{k}_i$ is the wave vector of the $i^{th}$ field. Plugging in the relation in Eq.~\ref{P_chi} and selecting appropriate terms corresponding to the DFWM signal,
\begin{align}
    P_{DFWM} = &\chi^{(3)}_0 E_1(\omega)E_2^*(\omega)E_3(\omega)e^{i(\bm{k}_1-\bm{k}_2+\bm{k}_3)\cdot \bm{r}} \notag \\
    &+ \chi^{(5)}_0 (|E_1|^2 + |E_2|^2 + |E_3|^2) E_1(\omega)E_2^*(\omega)E_3(\omega)e^{i(\bm{k}_1-\bm{k}_2+\bm{k}_3)\cdot \bm{r}} + \cdots
\end{align}
By substituting this expression into the wave equation, it can be shown that the DFWM signal is generated when the phase matching condition $\bm{k}_s = \bm{k}_1-\bm{k}_2+\bm{k}_3$ is satisfied as the three waves propagate through a finite-size nonlinear medium \cite{Boyd2008}. However, the polarization shows no explicit $e^{-i\omega \tau}$ dependence, indicating that higher-order ($\chi_0^{(5)}$) wave mixing alone cannot account for the oscillatory behavior observed in the experiment. To explain the oscillations in amplitude and chirp, we model the nonlinear susceptibility as a function of the net electric field $\chi^{(3)} \rightarrow \chi^{(3)}(\mathcal{E})$, rather than treating it as a constant. This field dependence arises from strong light-matter interaction, which dynamically changes the material's band structure during wave mixing.  We can expand the effective susceptibility $\chi^{(3)}(\mathcal{E})$ as the power series of the net field $\mathcal{E}$ as
\begin{align}
    P^{(3)}_{eff} = \left[\chi^{(3)}_0 + a\left(E_1(\omega)e^{-i\omega\tau} + E_2(\omega)e^{-i\omega\tau} + E_3(\omega)+c.c.\right) + \cdots \right]E_1(\omega)E_2^*(\omega)E_3(\omega)
\end{align}
where $\chi^{(3)}_0$ is the unperturbed susceptibility.  Expanding this relation gives,

\begin{align}
\label{P3_w}
     P^{(3)}_{eff} = f_0(\omega) + f_1(\omega)e^{-i\omega\tau} + g_1(\omega)e^{i\omega\tau} + f_2(\omega)e^{-2i\omega\tau} + g_2(\omega)e^{2i\omega\tau} + \cdots
\end{align}

Upon inverse transformation, the time-dependent third-order polarization will have terms that depend on $\cos{\omega\tau}$ due to the phase $e^{-i\omega\tau}$ in the spectrum. This means the amplitude of the DFWM output signal oscillates with the time delay, as is evident in the experiment.

Now, similarly, we can calculate the spectral phase of the third-order polarization as,
\begin{align}
    \phi(\omega) &= \arctan{\left(\frac{\Im(P^{(3)})}{\Re(P^{(3)})}\right)}\\
    &= \arctan{\left(\frac{\substack{\Im(f_0) + (\Im(f_1) + \Im(g_1))\cos{\omega\tau} - (\Re(f_1) - \Re(g_1))\sin{\omega\tau} \\
    + (\Im(f_2) + \Im(g_2))\cos{2\omega\tau} - (\Re(f_2) - \Re(g_2))\sin{2\omega\tau}}}{\substack{\Re(f_0) + (\Re(f_1) + \Re(g_1))\cos{\omega\tau}  (\Im(f_1) - \Im(g_1))\sin{\omega\tau} \\
    + (\Re(f_2) + \Re(g_2))\cos{2\omega\tau} + (\Im(f_2) - \Im(g_2))\sin{2\omega\tau}}}\right)}
\end{align}
using the approximation $\arctan(\theta) \approx \theta$ and Taylor expansion of $\cos\omega\tau$ and $\sin \omega\tau$ gives,
\begin{align}
    \phi(\omega) \approx \phi_0 + \alpha \omega + \beta \omega^2 + \cdots
\end{align}
$\beta$ is the coefficient of the $\omega^2$ term in the expansion and is called the spectral chirp.
\begin{align}
    \beta = -\frac{(\Im(f_1) + \Im(g_1))/2 + 2(\Im(f_2) + \Im(g_2)}{\substack{\Re(f_0) + (\Re(f_1) + \Re(g_1))\cos{\omega\tau} + (\Im(f_1) - \Im(g_1))\sin{\omega\tau} \\
    + (\Re(f_2) + \Re(g_2))\cos{2\omega\tau} + (\Im(f_2) - \Im(g_2))\sin{2\omega\tau}}}\tau^2
\end{align}
In general, the relation between temporal and spectral chirp is non-trivial, but within the narrow pulse limit, the temporal chirp is proportional to the reciprocal of the spectral chirp.
\begin{align}\label{temporal_chirp_sus_mod}
    \beta_t = \frac{1}{4\beta} = \frac{-1}{4\tau^2}\frac{\substack{\Re(f_0) + (\Re(f_1) + \Re(g_1))\cos{\omega\tau} + (\Im(f_1) - \Im(g_1))\sin{\omega\tau} \\
    + (\Re(f_2) + \Re(g_2))\cos{2\omega\tau} + (\Im(f_2) - \Im(g_2))\sin{2\omega\tau}}}{(\Im(f_1) + \Im(g_1))/2 + 2(\Im(f_2) + \Im(g_2)}
\end{align}
This shows that the temporal chirp not only has the $\cos\omega\tau$ terms, which means the temporal chirp also oscillates with fundamental/central frequency, but also contains higher frequency contributions due to $\cos{2\omega\tau}$ terms.

\subsection{Time-dependent perturbation theory with band-structure modulation}

In the DFWM process, the electron transitions from the Valence band (VB) to the Conduction band (CB) in a sufficiently strong incoming electric field. The output electric field $E_{DFWM} \propto \frac{\partial \langle J(t) \rangle^{(3)}}{ \partial t}$.

A simple extension of the Kubo formalism gives us a general expression of the third-order expectation value of current \cite{Bruus2004},
\begin{eqnarray}
\label{3rd_orde_kubo}
\langle J \rangle^{(3)} = (-i)^3 && \int_{t_0}^{t} dt_3 \int_{t_0}^{t_3} dt_2 \int_{t_0}^{t_2} dt_1
\Bigg\langle \bigg[\Big[ \big[\hat{J}(t), \hat{V}(t_1)\big], \hat{V}(t_2)\Big], \hat{V}(t_3) \bigg] \Bigg\rangle_0
\end{eqnarray}
where $V(t)$ is the external potential and $\hat{O}$ represents any operator $O$ in the interaction picture. The expectation value $\langle \rangle_0$ is with respect to the unperturbed Hamiltonian $H_0$. The external potential in the DFWM process is $V(t) = \int d^3\mathbf{r}\  \mathbf{J}(\mathbf{r}) \cdot \mathbf{A}_{ext}(\mathbf{r},t)$, with $\mathbf{J}(\mathbf{r})$ being the current operator and $\mathbf{A}_{ext}$ being the vector potential of the total incoming electric field $\mathbf{E}_{ext}(\mathbf{r},t) = \mathbf{E}_g(t -\tau) e^{i\omega_0 (t-\tau)-i\mathbf{k}_1\cdot \mathbf{r}} + \mathbf{E}_g(t-\tau) e^{i\omega_0(t-\tau)-i\mathbf{k}_2\cdot \mathbf{r}} + \mathbf{E}_p(t) e^{i\omega_0t - i\mathbf{k}_3\cdot \mathbf{r}} + c.c.$, where $\tau$ is the time delay between Probe and Gate pulses. Plugging all the terms into Eq~\ref{3rd_orde_kubo},
\begin{align}
    \label{full_J3_eq}
    \langle J^i (t)\rangle^{(3)} = \frac{(-i)^3}{\sqrt{A}} &\int_{t_0}^{t} dt_3 \int_{t_0}^{t_3} dt_2 \int_{t_0}^{t_2} dt_1\ A_{ext}^{\alpha}(t_1) A_{ext}^{\beta}(t_2) A_{ext}^{\gamma}(t_3) \notag \\
    &\times \underbrace{j^i_{m,n,\mathbf{q}} j^{\alpha}_{m_1,n_1,\mathbf{q}_1} j^{\beta}_{m_2,n_2,\mathbf{q}_2} j^{\gamma}_{m_3,n_3,\mathbf{q}_3}}_{\text{time-independent terms}}\notag\\
    & \times \exp{\left(i\Omega_{m,n,\mathbf{q}} t + i\Omega_{m_1,n_1,\mathbf{q}_1} t_1 + i\Omega_{m_2,n_2,\mathbf{q}_2} t_2 + i\Omega_{m_3,n_3,\mathbf{q}_3} t_3 \right)}\notag\\
    \times &\underbrace{\Bigg\langle \bigg[\Big[ \big[a^{\dagger}_{m,\mathbf{q}} a_{n,\mathbf{q}}, a^{\dagger}_{m_1,\mathbf{q}_1} a_{n_1,\mathbf{q}_1}\big], a^{\dagger}_{m_2,\mathbf{q}_2} a_{n_2,\mathbf{q}_2}\Big],a^{\dagger}_{m_3,\mathbf{q}_3} a_{n_3,\mathbf{q}_3} \bigg] \Bigg\rangle_0}_{=\mathcal{O}(\text{Operator part})}
\end{align}
Here, $\Omega_{m,n,\mathbf{q}}$ represents the energy difference between the $m^{\text{th}}$ and $n^{\text{th}}$ bands and $t_0$ is the initial time when there is no interaction\footnote{In most cases, $t_0 = -\infty$ indicating no perturbation at time $-\infty$}. For brevity, Einstein's summation convention is applied to all indices in the last expression. For the remainder of the calculation, we assume that the momentum carried by the photons ($\sim 2\pi/\lambda$) is an order of magnitude smaller than the crystal momentum of the electrons ($\sim 2\pi/a$, $a$ is the lattice parameter of MgO). Atomic units are used throughout these derivations. Thus, photon-assisted electron excitation results in a negligible change in crystal momentum $\mathbf{k}$\footnote{The change would be significant if we included the effects of phonons, which are not considered in this model.}. The $j^i_{m,n,\mathbf{q}}$ represents the matrix elements of the current operators and is proportional to the transition dipole matrix elements between the valence band (VB) and the conduction band (CB). Since the current matrix element (transition-dipole moment) varies weakly as a function of momentum $k$ \cite{Vampa2014, Hohenleutner2015}, it can be considered a constant and absorbed into the normalization constant. The electronic band structure of MgO was calculated using density functional theory (DFT) within the generalized gradient approximation (GGA), employing the Perdew–Burke–Ernzerhof (PBE) functional for exchange–correlation. It is well established that the PBE functional tends to underestimate the band gap in wide-bandgap insulators \cite{Sham1983, Crowley2016, Liu2024}. Therefore, the calculated band gap of approximately $5~\mathrm{eV}$ was corrected to the experimental value of $7.8~\mathrm{eV}$ using the scissors approximation \cite{Levine1989}.

The operator part in Eq.~\ref{full_J3_eq} can be solved using the anti-commutation relation of fermions. The appropriate non-zero terms that satisfy the time ordering condition $t_1<t_2<t_3$ are chosen. In general, converging all orders of the response function requires considering multiple conduction bands \cite{gorelova2024}. Since we are only interested in the third-order response, it turns out that only two bands are sufficient to describe the off-resonant DFWM interaction in MgO ($\hbar \omega_0 = 1.56eV < \Delta_{\text{gap}}=7.78eV$). In addition, the intensity of the IR pulses in the experiment is not large enough for the higher-order multiphoton absorption process to contribute to the lower-order response function\cite{gorelova2024}. Thus, $\mathcal{O} \sim \langle a_m^{\dagger} a_m \rangle - \langle a_n^{\dagger} a_n \rangle \approx 1$ if $m \neq n$ for a two-band system.

\begin{figure}[t]
    \centering
    \includegraphics[width=\linewidth]{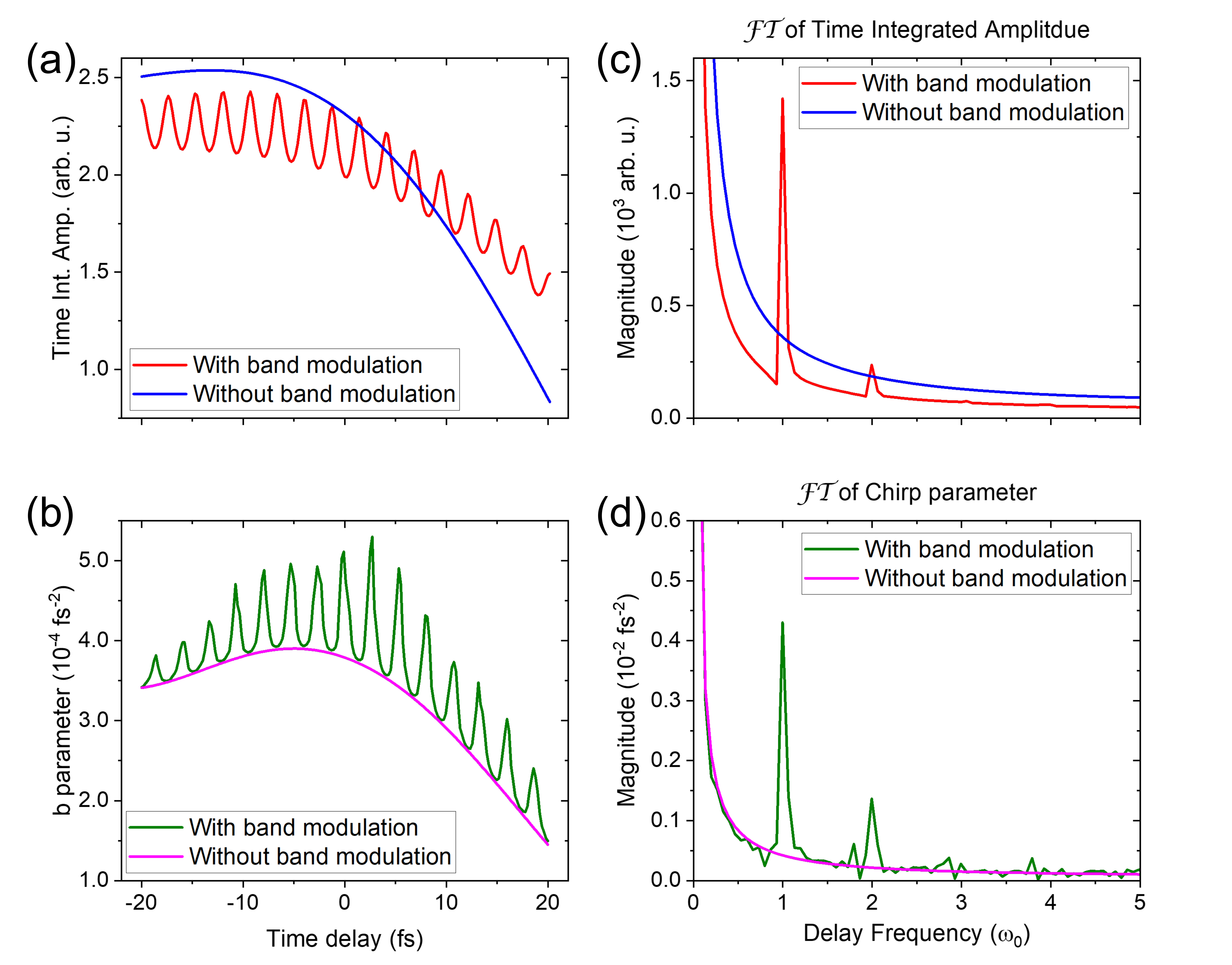}
    \caption{(a) Time-integrated amplitude and (b) temporal chirp are calculated using time-dependent perturbation theory with and without band structure modulation. Their corresponding Fourier transforms in the time-delay domain are shown in (c) and (d), respectively.}
    \label{fig:band_vs_noband}
\end{figure}

According to the Bloch acceleration theorem, the single-particle states become time dependent in reciprocal space. The crystal momentum transforms as $\mathbf{p}(t) = \mathbf{p}_0 - e\int^t_{t_0} \mathbf{E}_{ext}(t') dt'$. Thus, the valence and conduction band energies will also be modified as $\epsilon_{i}(t) = \epsilon_{i}\left(\mathbf{p}_0 - e\int^t_{t_0} \mathbf{E}_{ext}(t') dt'\right)$ with $i = \{CB, VB\}$.

Eq~\ref{full_J3_eq} is then simplified to get,
\begin{eqnarray}
    \label{new_J3_eq}
    \langle J^i (t)\rangle^{(3)} = &&\mathcal{N}^{i\alpha\beta\gamma}_{\mathbf{q}} \int_{t_0}^{t} dt_3 \int_{t_0}^{t_3} dt_2 \int_{t_0}^{t_2} dt_1\ A_{ext}^{\alpha}(t_1) A_{ext}^{\beta}(t_2) A_{ext}^{\gamma}(t_3) \notag\\
        &&\times \exp\left(i\int_{t_0}^{t}\Theta_{0,1,\mathbf{q}}(t') dt' + i\int_{t_0}^{t_1}\Theta_{1,0,\mathbf{q}}(t')dt' \right. + \left. i \int_{t_0}^{t_2}\Theta_{0,1,\mathbf{q}}(t')dt' + i\int_{t_0}^{t_3}\Theta_{1,0,\mathbf{q}}(t')dt' \right)
\end{eqnarray}
where, $\Theta_{1,0, \mathbf{q}}(t)$ is the time-dependent energy gap and is defined by $\Theta_{1, 0, \mathbf{q}}(t) = \epsilon_{CB}(t) - \epsilon_{VB}(t)$. Also, $ 0$ and $ 1$ correspond to $ VB$ and $ CB$, respectively, in the two-band model.

The $E_g(t)$ and $E_p(t)$ are the envelope functions for the Gate and Probe pulses, which are adopted from the experimental data \cite{walz2022}. Since the Keldysh parameter for our experiment is large ($>\sim 2.5$), the band modification is much smaller than in the case of a small Keldysh parameter\cite{Neufeld2022, gruzdev2018}. Thus, for numerical convenience, we assume the parabolic band approximation, i.e. $\epsilon_{VB} \approx -|\mathbf{p}|^2/2m_{VB}$ and $\epsilon_{CB} \approx \Delta\left(1 + |\mathbf{p}|^2/(2m_{CB}\Delta)\right)$. The effective masses $m_{VB}$ and $m_{CB}$ are extracted by fitting to the calculated band structure.

Lastly, the signal was measured along the direction of $\mathbf{k}_1 - \mathbf{k}_2 + \mathbf{k}_3$ in the experiment. Since $\mathbf{A}_{ext}(t)$ is composed of six terms (3 + 3 cc), Eq.~\ref{new_J3_eq} will have contributions from 216 terms. Of these, only 12 (6 + 6 cc) of the terms contribute to the phase matching condition of $\mathbf{k}_1 - \mathbf{k}_2 + \mathbf{k}_3$.

In Fig.~\ref{fig:band_vs_noband} (a) and (b), we show that the time-integrated amplitude and temporal chirp parameter respectively, calculated using time-dependent perturbation theory without the inclusion of band modulation (blue/magenta curves, see Eq. (8) of the main text) and after explicitly including the band structure modulation (red/green curves, see Eq. (9) of the main text). Fig.~\ref{fig:band_vs_noband} (c) and (d) are the Fourier transformation of (a) and (b), respectively. This shows that the peaks we see in the experimentally measured field observables in Fig. \ref{fig: FFT_intensity_chirp} are due to the band structure modulation of MgO as it interacts with the Gate and Probe pulses.

\subsection{Semiconductor Bloch Equation}
To fully understand strong field–matter interactions, we employ a non-perturbative microscopic model based on a one-dimensional two-band representation of MgO. When the crystal interacts with a strong laser field $E(t)$, electrons are driven from the valence band to the conduction band, the transition strength being determined by the element of the transition dipole matrix
\begin{equation}
d^{cv}_k = \frac{1}{i a_0}\int u^*_{c,k}\frac{\partial}{\partial k}u_{v,k} dx,
\end{equation}
where $u_{n,k}$ are the cell-periodic parts of the Bloch functions and $a_0$ is the lattice constant \cite{Wu2015}.

The transient electron dynamics under a strong, time-dependent electric field are described using the semiconductor Bloch equations (SBEs), which govern the evolution of the electron (hole) populations $n^e_k$ ($n^h_k$) and the microscopic polarization $p^{h,e}_k$ between them \cite{Schubert2014, Hohenleutner2015, Huttner2017}:
\begin{subequations}
\begin{align}
\hbar\frac{\partial n^e_k}{\partial t} &= -2\hbar \Im\left[d^{cv}k E(t)\right] + |e|E(t)\nabla_k n^e_k - \frac{1}{2T_1}(n^e_k - n^e{-k}),\\
\hbar\frac{\partial n^h_k}{\partial t} &= -2\hbar \Im\left[d^{cv*}k E(t)\right] + |e|E(t)\nabla_k n^h_k - \frac{1}{2T_1}(n^h_k - n^h{-k}),\\
i\hbar\frac{\partial p^{h,e}_k}{\partial t} &= \left(\epsilon^e_k + \epsilon^h_k - i\frac{\hbar}{T_2}\right)p^{h,e}_k - d^{cv}_k E(t)\left(1 - n^e_k - n^h_k\right) + i|e|E(t)\nabla_k p^{h,e}_k.
\end{align}
\end{subequations}
Here, $E^e_k$ and $E^h_k$ denote the electron and hole dispersion relations, respectively, defined as $E^e_k = E^c_k$ and $E^h_k = -E^v_k$, where $E^c_k$ and $E^v_k$ correspond to the conduction and valence band energies. The phenomenological relaxation and dephasing times are set to $T_1 = 7~\mathrm{fs}$ and $T_2 = 1.1~  \mathrm{fs}$ \cite{Hohenleutner2015, Huttner2017}. The input electric field parameters corresponding to the experiment are used in these calculations. From the time-dependent evolution of the electron (hole) populations and the microscopic polarization, we calculate the intraband and interband currents, which serve as the primary current sources arising from the acceleration of excited carrier wavepackets within the bands and the temporal variation of the polarization between electron–hole pairs, respectively \cite{Huttner2017}.
\begin{subequations}
\begin{align}
J_{\mathrm{intra}}(t) = \frac{e}{\hbar}\sum_{k,i}\nabla_{k}\epsilon^i_{k}n^{i}_k\\
J_{\mathrm{inter}}(t) = \sum_{k}d^{cv}_k p^{h,e}_k
\end{align}
\end{subequations}
where $i=\{e,h\}$. Both current will contribute to the total ouput electric field $E_{\mathrm{out}}(t)\propto J_{\mathrm{intra}}(t) + \frac{\partial}{\partial t}J_{\mathrm{inter}}(t)$.

\end{document}